\documentclass[aps,pre,twocolumn,longbibliography]{revtex4-1}
\usepackage{amsfonts,amssymb,amsmath,latexsym,epsfig}
\usepackage{xcolor}
\usepackage{graphics}
\usepackage{epsfig}
\usepackage{color}
\newcommand{\red}{\textcolor{black}}

\newcommand{\be}{\begin{equation}} 
\newcommand{\ee}{\end{equation}}
\newcommand{\bea}{\begin{eqnarray}} 
\newcommand{\eea}{\end{eqnarray}}

\begin{document}
%%\begin{CJK*}{GBK}{}

\title{Coexistence of fast and slow gamma oscillations \\ in one population of 
inhibitory spiking neurons}
\author{Hongjie Bi$^{1}$}
\email{hongjie.bi@u-cergy.fr}
\author{Marco Segneri$^{1}$}
\email{marco.segneri@u-cergy.fr}
\author{Matteo di Volo$^{1,2}$}
\email{matteo.divolo@unic.cnrs-gif.fr}
\author{Alessandro Torcini$^{1}$}
\email{alessandro.torcini@u-cergy.fr}

\affiliation{ $^{1}$Laboratoire de Physique Th\'eorique et Mod\'elisation, Universit\'e de Cergy-Pontoise,
CNRS, UMR 8089, 95302 Cergy-Pontoise cedex, France\\
$^{2}$Unit\'e de Neuroscience, Information et Complexit\'e (UNIC),
CNRS FRE 3693, 1 avenue de la Terrasse, 91198 Gif sur Yvette, France}

\date{\today}

\begin{abstract}
Oscillations are a hallmark of neural population activity
in various brain regions with a spectrum covering a wide range of frequencies.
Within this spectrum gamma oscillations have received particular attention due to their 
ubiquitous nature and to their correlation with higher brain functions.
Recently, it has been reported that gamma oscillations in the hippocampus of behaving rodents
are segregated in two distinct frequency bands: slow and fast. 
\red{These two gamma rhythms correspond to dfferent states of the network, 
but their origin has been not yet clarified.} Here, we show theoretically and numerically 
that a single inhibitory population can give rise to coexisting slow and fast gamma rhythms 
corresponding to collective oscillations of a balanced spiking network.
The slow and fast gamma rhythms are generated via 
two different mechanisms: the fast one being driven by the coordinated tonic neural firing
and the slow one by endogenous fluctuations due to irregular neural activity.
We show that almost instantaneous stimulations can switch the collective gamma oscillations 
from slow to fast and vice versa. Furthermore, to make a closer contact with 
the experimental observations, we consider the modulation of the gamma rhythms induced 
by a slower (theta) rhythm driving the network dynamics. \red{In this context, 
depending on the strength of the forcing, we observe phase-amplitude and 
phase-phase coupling between the fast and slow gamma oscillations and the theta forcing.
Phase-phase coupling reveals different theta-phases preferences for the two coexisting
gamma rhythms.}
\end{abstract}


\maketitle
%%\end{CJK*}

\section{Introduction}
The emergence of collective oscillations in complex system
has been a subject largely studied in the last decades 
from an experimental as well as from a theoretical point of view,
for a recent review see \cite{pikovsky2015}. In particular,
the transition from asynchronous to collective dynamics
in heterogeneous oscillators networks
has been characterized in terms of methods borrowed
from statistical mechanics \cite{winfree,kuramoto2012,hong2007}
and nonlinear dynamics \cite{crawford1994,strogatz2000,barre2016}. 
Exact analytic techniques to reduce the infinite dimensional dynamics 
of globally coupled inhomogeneous phase oscillators to few 
mean field variables have became available in the last decade \cite{ott2008}
allowing for noticeable progresses in the field \cite{pikovsky2015}.
Quite recently, these reduction techniques have been 
applied to globally coupled spiking neural networks \cite{montbrio2015},
thus opening new perspectives for the study of large
ensembles of spiking neurons and for the understanding
of the mechanisms underlying brain rhythms.

Oscillatory dynamics is fundamental for the functioning of the mammalian brains,
rhythms ranging from 1 to 500 Hz have been measured at a mesoscopic level, corresponding to the dynamics of neural populations, by employing electroencephalography (EEG), magnetoencephalography (MEG), or local field potential (LFP) \cite{buzsaki2006}.

In particular, gamma oscillations (30-100 Hz) have been
suggested to underlie various cognitive and motor functions.
Oscillations in the gamma band have been related to
attention selection \cite{fries2001}, memory formation and retrieval \cite{bragin1995,fell2001}, 
binding mechanisms for sensory awareness \cite{engel2001},
and human focal seizures \cite{truccolo2014}.

Gamma oscillations have been observed in many areas of the brain and their emergence has been shown to be crucially dependent on inhibitory networks \cite{buzsaki2012,bartos2007synaptic}.
By following  \cite{buzsaki2012} gamma oscillations in purely inhibitory
networks can emerge only via two mechanisms: the single neurons can 
fire periodically locked in phase \cite{kopell2002} or each neuron
can have irregular activity, but sufficiently strong recurrent
interactions can render the asynchronous state unstable against
fluctuations and collective oscillations (COs) can arise \cite{brunel1999,brunel2000,matteo}.
\red{On one hand, the role of the synaptic mechanisms in promoting tonic synchronization
in the gamma range has been clarified in \cite{vida2006,bartos2007synaptic}.
On another hand, fast network oscillations with irregular neural
discharges can emerge  when the neurons are
operating in the so-called balanced state \cite{bal1,bal2,bal3,rosenbaum2014,bal4}.}
A typical cortical state, where the balance of excitation and inhibition
allows for a healthy activity of the brain. The balanced state has
been observed {\it in vitro} and {\it in vivo} experiments in the
cerebral cortex \cite{berg2007,barral2016} and reported in 
simulations of networks of excitatory and inhibitory spiking neurons \cite{brunel2000,brunel2003,wolf} as well 
as of purely inhibitory circuits driven by external excitatory currents \cite{jahnke2008,montefortePRX}.

Gamma oscillations are usually modulated by theta oscillations  
in the hippocampus during locomotory actions and rapid eye movement (REM) sleep, theta frequencies correspond to 4-12 Hz in rodents  \cite{buzsaki2002,belluscio2012} and to 1-4 Hz in humans \cite{jacobs2014,zhang2015}.
Two mechanisms of entrainment (or cross-frequency coupling) between theta and gamma oscillations have been reported : 
namely, phase-amplitude (P-A) and phase-phase (P-P) coupling. The P-A coupling (or theta-nested 
gamma oscillations) corresponds to the fact that the phase of the theta-oscillation modifies the amplitude of the
gamma waves \cite{white2000,butler2016}, while P-P coupling refers to n:m phase locking between gamma and theta phase oscillations \cite{tass1998,belluscio2012}.

Recently, the co-existence of gamma oscillations
in three distinct bands has been reported for the cornu ammonis area 1 (CA1) of the hippocampus \cite{belluscio2012}:
\red{namely, a slow one ($\simeq$ 30-50 Hz), a fast (or intermediate) one ($\simeq$ 50-90 Hz), 
and a so called $\varepsilon$-band ($\simeq$ 90-150 Hz). However, only the two lower bands show a clear correlation (P-P coupling) with  the theta rhythm during maze exploration and REM sleep, thus suggesting their functional relevance \cite{belluscio2012}.
There are several further evidences that these two gamma bands
correspond to different states of the hippocampal network \cite{zheng2015}.
In particular, in freely behaving rats place cells code differently the space location and the running speed 
during theta-nested slow or fast gamma rhythms \cite{bieri2014,zheng2015,zheng2016}.
Moreover, gamma  rhythms with similar low and high frequencies subtypes occur in many other
brain regions, besides the hippocampus \cite{sirota2008,colgin2016}.
Despite their relevance, the mechanisms behind the
emergence of these two distinct gamma bands are not yet clarified.}

\red{For what concerns the hippocampus, experimental evidences show that slow gamma rhythms couple the activity
of the CA1 area to synaptic inputs from CA3, while fast gamma rhythms in CA1 are entrained by
inputs from medial Entorinhal Cortex (mEC) \cite{colgin2016}. Slow and fast oscillations
have been recorded also in CA3, where fast gamma are entrained by synaptic inputs from mEC \cite{colgin2009}.
These findings suggest that CA3-activated interneurons drive slow gamma, while mEC-activated interneurons drive fast gamma.  Nonetheless, it has been shown that a substantial proportion of CA1 interneurons phase-lock to both slow and fast gamma LFP oscillations \cite{colgin2009,belluscio2012, schomburg2014}.
Therefore, as suggested by L.L. Colgin in \cite{colgin2016}, such interneurons may be part of
a network that can generate either slow or fast gamma,
depending on the state of the network. 
Furthermore there are clear evidences that gamma rhythms can be generated locally {\it in vitro}
in the CA1, as well as in the CA3 and mEC, thanks to optogenetic
stimulations \cite{akam2012,pastoll2013,butler2016} or 
pharmacological manipulations, but at lower gamma frequencies
with respect to optogenetics \cite{fisahn1998,traub2003,pietersen2014,craig2015}.} 
\red{A recent theoretical work has analysed the emergence
of gamma oscillations in a neural circuit composed by two populations of interneurons 
with fast and slow synaptic time scales \cite{keeley2016}. 
Based on the results of this idealized rate model and on the analysis of experimental data sets
for the CA1 area the authors showed that multiple gamma bands can arise locally 
without being the reflection of feedfoward inputs.}

In the present work, we show, for the first time to our knowledge, that a single inhibitory population, characterized by only one synaptic time,
can display coexisting fast and slow gamma COs \red{corresponding to different network states.}
In particular, the slow gamma oscillations are associated to 
irregular spiking behaviours and are fluctuations driven, while
the fast gamma oscillations coexist with a much more regular 
neural dynamics and they can be characterized as mean driven \cite{renart2007,angulo2017}.
Furthermore, in presence of theta forcing \red{we observe different 
theta-gamma cross-frequency coupling scenarios depending on the forcing
amplitude. For small amplitudes we have theta-nested gamma oscillations
resembling those reported for various brain areas
{\it in vitro} under optogenetic sinusoidal theta-stimulation \cite{akam2012,pastoll2013,butler2016}.
At larger amplitudes the two types of gamma COs phase lock to the theta rhythm, 
similarly to what has been reported experimentally for the CA1 region of the hippocampus \cite{colgin2009, belluscio2012}.}
More specifically we have studied balanced sparse inhibitory
networks of quadratic integrate-and-fire (QIF) neurons 
pulse coupled via inhibitory post-synaptic potentials (IPSPs),
characterized by a finite synaptic time scale.
For this sparse network we derived an effective mean-field (MF) 
by employing recently developed reduction techniques for QIF networks~\cite{montbrio2015,devalle2017,coombes2019,matteo}.
\red{In the MF model, in proximity of the sub-critical Hopf bifurcations,
we report regions of bistability involving one stable focus and one stable 
limit cycle.} In direct simulations of the corresponding spiking network
we observe the coexistence of two distinct COs
with frequencies in the slow and fast gamma band.
The slow gamma COs are due to the
microscopic irregular dynamics, characteristic of the balanced dynamics, 
which turns the damped oscillations towards the MF focus in
sustained COs. The fast gamma COs are instead
related to the oscillatory branch emerging via the sub-critical Hopf bifurcation from
the asynchronous state. The network can be driven from one kind of COs to the other 
by transiently stimulating the neurons. \red{In presence of a theta forcing
nested gamma oscillations characterized by a P-A coupling appear for small forcing amplitudes,
while at intermediate amplitudes slow and fast gamma 
phases lock to the theta phase displaying P-P coupling between the rhythms.
For even larger amplitudes only fast gamma are observables with a maximal power
in correspondence of the maximum of the stimulation.}

The paper is organized as follows. In Sec. II, we introduce the model for an inhibitory sparse balanced network
of QIF neurons as well as the macroscopic and microscopic indicators employed to characterize its
dynamics. Section III is devoted to the derivation of the corresponding effective MF model and to
the linear stability analysis of the asynchronous state. Simulation results for the network for high and
low structural heterogeneity are reported in Section IV and compared with MF forecasts.
The coexistence and transitions from slow (fast) to fast (slow) gamma oscillations is analyzed in Section V 
together with the cross-frequency coupling between theta and gamma oscillations.
A concise discussion of the results and of possible future developments
is reported in Section VI. Finally, Appendix A is devoted to the analysis of
coexisting gamma oscillations in Erd\"os-Reniy networks, \red{while Appendix B
discusses of a general mechanism for the coexistence of noise-driven and tonic oscillations.}

\section{Methods}

\subsection{The network model}

We consider $N$ inhibitory pulse-coupled QIF neurons~\cite{ermentrout1986} arranged 
in a random sparse balanced network. The membrane potential of each neuron 
evolves according to the following equations:
\begin{subequations}\label{eq:1}
\begin{eqnarray}
&\tau_{m} \dot{v}_{i}(t) = I + v_{i}^2(t) 
- \tau_m J y_i(t)  
\label{eq:1a}
\\
& \tau_{d} \dot{y}_{i}(t) = -y_{i}(t)+ \sum_j\epsilon_{ji} \delta(t - t_{j}(m))
\enskip ,
\end{eqnarray}
\end{subequations}
where $\tau_{m}=15$ ms represents the membrane time constant,
$I$ an external DC current, encompassing the effect of distal
excitatory inputs and of the internal neural excitability.
\red{The last term in \eqref{eq:1a} is the inhibitory synaptic current,
with $J$ being the synaptic coupling and $y_i$ the synaptic field
seen by neuron $i$.} Whenever the membrane potential ${v}_{i}$ reaches infinity
a spike is emitted and ${v}_{i}$ resetted to $-\infty$.

The field $y_i$ is the linear super-position of all the 
\red{exponential IPSPs $s(t)= \exp{(-t/\tau_d)}$} received by the neuron $i$ from its
pre-synaptic neurons in the past, namely
\be
y_i (t) = \frac{1}{\tau_d} \sum_{j \in pre(i)} \sum_{m | t_j(m) < t} \epsilon_{ji} \Theta(t-t_j(m)) s(t-t_j(m))
\ee
\red{where $\tau_d$ is the synaptic time constant,} $t_{j}(m)$ the spike time of the 
$m$-th spike delivered by the $j$-th neuron, $\Theta(t)$ is the Heaviside function
and $\epsilon_{ji}$ is the adjacency matrix of the network. In particular,
$\epsilon_{ji} = 1$  (0) if a connection from node $j$ to $i$ exists (or not) and
$k_i=\sum_j\epsilon_{ji}$ is the number of pre-synaptic neurons connected
to neuron $i$, or in other terms its in-degree. 

In order to compare the simulation results with an exact MF 
recently derived \cite{montbrio2015,devalle2017,matteo}, we consider sparse 
networks where the in-degrees $k_i$ are extracted from a Lorentzian distribution 
\begin{eqnarray}\label{eq:2}
P(k)= \frac{\Delta_k}{(k-K)^2+\Delta_k^2}
\end{eqnarray}
peaked at $K$ and with a half-width half-maximum (HWHM) $\Delta_k$,
the parameter $\Delta_k$ measures the level of structural
heterogeneity in the network, and analogously to Erd\"os-Renyi networks 
we assumed the following scaling for the HWHM $\Delta_k=\Delta_0\sqrt{K}$.
\red{The DC current and the synaptic coupling are rescaled with the median in degree $K$ as
$I= I_0 \sqrt{K}$ and $J=J_0/\sqrt{K}$, as usually done to achieve 
a self-sustained balanced state for sufficiently large in degrees \cite{bal1,bal2,bal3,bal4,wolf}.} 
In this paper we will usually consider $I_0=0.25$, $N=10,000$ and $K=1,000$, unless stated
otherwise.

\subsection{Simulation Protocols}

\red{The network dynamics is integrated by employing a standard Euler scheme
with an integration time step $\Delta t = \tau_m/10000$. 
The coexistence of solutions in proximity of
a sub-critical Hopf bifurcation is analysed by performing adiabatic network simulations
where a control parameter (e.g. the synaptic time $\tau_d$) is slowly varied.
In particular, these are performed by starting with a initial value of 
$\tau_d^{(0)}$ and arriving to a final value $\tau_d^{(1)}$ in $M$ steps,
each time increasing $\tau_d$ by $\Delta \tau_d = (\tau_d^{(1)}-\tau_d^{(0)})/(M-1)$.
Once the final value $\tau_d^{(1)}$ is reached, the synaptic time is decreased in
steps $\Delta \tau_d$ down to $\tau_d^{(0)}$.
Each step corresponds to a simulation for a time $T_s= 90$ s during
which the quantities of interest are measured, after discarding a transient $T_t = 15$ s.
The initial condition for the system at each step is its final 
configuration at the previous step.}

\red{For what concerns the analysis of the crossing times $t_c$ from slow (fast) to fast (slow)
gamma in a bistable regime, reported in Section V A, we proceeded as follows.
Let us first consider the transition from slow to fast gamma COs.  We initialize the
system in the slow gamma state at a current $I_0 \equiv I_1$ ensuring the 
bistability of the dynamics. Then we increase the DC current to a value $I_0 \equiv I_2$ 
for a time interval $T_P$, after that time we return to the original value $I_0 \equiv I_1$
and we check, after a period of 1.5 s, if the system is in the slow or fast gamma regime.
Then we repeat the process $M=30$ times for each considered value of $T_P$
and we measure the corresponding transition probability. The crossing time $t_c$
is defined as the minimal $T_P$ giving 80 \% of probability that the transition will take place.
To analyse the transition from fast to slow, we initialize the system
in the fast gamma state at a DC current $I_1$, we decrease the current to a value $I_0 \equiv I_3$
for time $T_P$ and the we proceed as before.
To examine the influence of noise on such transitions we added to the membrane potential
evolution a noise term of zero average and amplitude $A_n$.
}

\subsection{Indicators}

To characterize the collective dynamics in the network we measure the 
mean membrane potential ${V}(t) = \sum_{i=1}^N v_i(t)/N$, the instantaneous firing 
rate $R(t)$, corresponding to the number
of spikes emitted per unit of time and per neuron, as well as 
the mean synaptic field ${Y}(t) = \sum_{i=1}^N y_i(t)/(NK)$ \cite{field}.
 
The microscopic activity can be analyzed by considering
the inter-spike interval (ISI) distribution as characterized by
the coefficient of variation $cv_i$ for each neuron $i$,
which is the ratio between the standard deviation and the mean of the 
ISIs associated to the train of spikes emitted by the considered neuron. 
In particular, we will characterize each network in terms of the average coefficient of
variation defined as $CV = \sum_i cv_i/N$.
Time averages and fluctuations are usually estimated on time intervals 
$T_s \simeq 90$ s, after discarding transients $T_t \simeq 15$ s.

\red{Phase entrainement between an external forcing characterized by its phase $\theta(t)$
and the collective oscillations induced in the network can be examined by considering the following phase difference:
\begin{equation}
 \Delta_{nm}(t) = n*\theta(t) - m*\gamma(t) \enskip;
\label{phase}
\end{equation}
where $\gamma(t)$ is the phase of the COs defined by considering the time occurrences $T_k$ of the $k$ maximum of the instantaneous firing rate $R(t)$ of the network,
namely $\gamma(t) = 2 \pi (t-T_k)/(T_{k+1}-T_k)$ with $t \in [T_k, T_{k+1}]$ \cite{phase}.
We have a $n:m$ phase locking whenever the phase difference \eqref{phase} is bounded
during the time evolution, i.e. $|\Delta_{nm}(t)| < const$.}
 
\red{This somehow qualitative criterion can be made more quantitative by considering  
statistical indicators measuring the level of $n:m$ synchronization
for irregular/noisy data. In particular, an indicator based on the
Shannon entropy has been introduced in \cite{tass1998}, namely 
\begin{equation}
e_{nm}=\frac{(E_{max}-E)}{E_{max}}   \quad {\rm with}  \quad E= - \sum_{k=1}^M p_k \ln(p_k)
\label{entropy}
\end{equation}
where $E$ is the entropy associated to the
distribution of $\Delta_{nm}(t)$ and $E_{max}= \ln(M)$ with $M$ number of bins. }

\red{The degree of synchronization among the phases can be also measured 
by the so-called Kuramoto order parameter, namely
\cite{kura,belluscio2012}
\begin{equation}
\rho_{nm} = \left| \frac{1}{L} \sum_{k=1}^L {\rm e}^{i \Delta_{nm}(t_k)} \right| \quad ;
\label{kura}
\end{equation}
where $| \cdot |$ represents the modulus and  $t_k=k \frac{T_W}{L}$ are $L$ successive 
equispaced times within the considered time window $T_W$. For completely desynchronized 
phases $\rho_{nm} \propto 1/\sqrt{L}$,
while partial (full) synchronization will be observable whenever $\rho_{nm}$ is finite (one).
}

\red{To assess the stationarity and the statistical significance of the obtained
data we measured the above indicators within a time window $T_W$ and we averaged
the results over several distinct time windows in order to obtain also the corresponding 
error bars. Furthermore, to avoid the detection of spurious phase locking due to noise or band-pass
filtering one should derive significance levels $e^{(S)}_{nm}$ and $\rho^{(S)}_{nm}$
for each $n:m$ phase locking indicators $e_{nm}$ and $\rho_{nm}$ \cite{tass1998,scheffer2016}. 
The significance levels have beeen estimated by considering surrogate data obtained by randomly shuffling
the original time stamps of one of the two considered phases.
Moreover, by following \cite{scheffer2016} we considered also other two types
of surrogates for the generation of $\Delta_{nm}(t)$ \eqref{phase} within a certain time window $T_W$. 
These are the time-shift surrogate, obtained by time shifting the origin of one time series for the phases with respect to the original one in the definiton of \eqref{phase} and the random permutation
surrogate, obtained by randomly choosing the origins of two time windows of duration $T_W$ to estimate 
$\Delta_{nm}(t)$.  
}

\section{Effective mean-field model for a sparse QIF network}

By following \cite{matteo} we derive an
effective MF formulation for the model \eqref{eq:1}.
As a starting point we consider an exact macroscopic model recently derived
for fully coupled networks of pulse-coupled QIF \cite{montbrio2015}, in particular
we focus on inhibitory neurons coupled via exponentially decaying IPSPs \cite{devalle2017}.
For a \red{structurally inhomogeneous network made of identical QIF neurons,
with the synaptic couplings randomly distributed according to a Lorentzian,
the MF dynamics can be expressed}
in terms of only three collective variables (namely, $V$,$R$ and $Y$), as follows : 
\begin{subequations}\label{mf_0}
\bea
\tau_m \dot{R} &=&  2 R V + \frac{\Gamma}{\pi} Y \\
\tau_m \dot{V} &=& {V^2 + I} + {\bar g}   \tau_m Y -(\pi \tau_m R)^2 \\
\tau_d \dot{Y} &=& - Y + R
\eea
\end{subequations}
where ${\bar g}$ is the median and $\Gamma$ the HWHM of the Lorentzian distribution of the synaptic couplings.

\begin{figure}
\centerline{\includegraphics[scale=0.45]{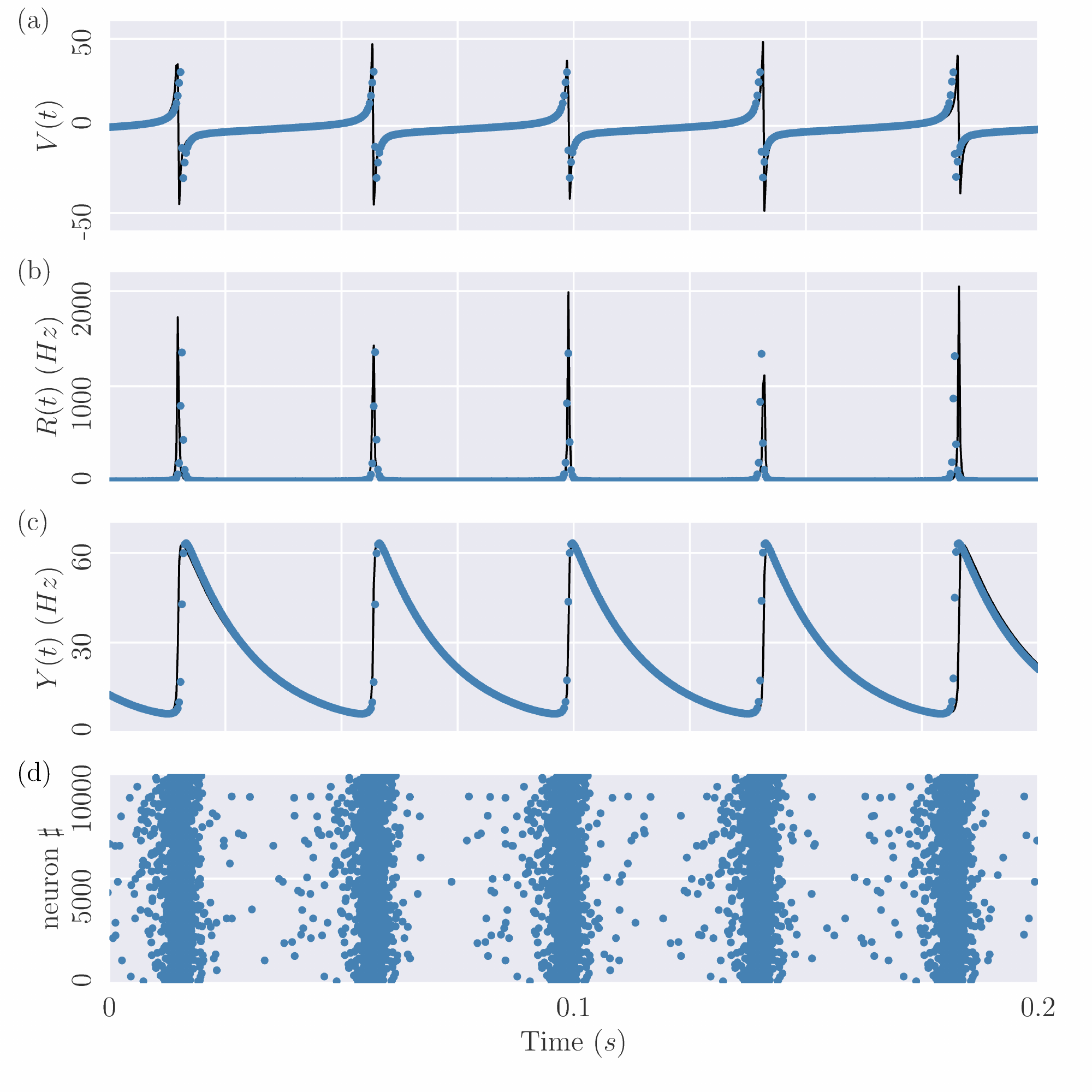}}
\caption{{\bf Comparison of the spiking dynamics with the mean-field results.}
Collective variables $V$ (a), $R$ (b) and $Y$ (c) versus time,
obtained from simulations of the spiking network \eqref{eq:1} (blue circles)
as well as from the MF formulation \eqref{eq:5a} (black line).
In (d) the corresponding raster plot is also displayed, revealing
clear COs with frequency $\nu_{OSC} \simeq 24$ Hz.
Dynamics of the network of $N = 10000$ neurons with median in-degree $K = 1000$ and $\Delta_0 = 0.3$.  
Other parameters are $I_0 = 0.25$, $J_0=1.0$ and $\tau_d=15$ ms.}
\label{f0}
\end{figure}

\red{At a mean-field level, the above formulation can be applied to a sparse network,
indeed the quenched disorder in the connectivity distribution
can be rephrased in terms of a random synaptic coupling. Namely, each neuron $i$ is subject in average to an  inhibitory synaptic current of amplitude $g_0 k_i Y/(\sqrt{K})$ proportional to its in-degree $k_i$.
Therefore at a first level of approximation} we can consider the neurons as fully coupled, but with
random values of the coupling distributed as a 
Lorentzian  of median ${\bar g} = - J_0 \sqrt{K}$ and HWHM $ \Gamma = J_0 \Delta_0$. 
The MF formulation \eqref{mf_0} takes now the expression:
\begin{subequations}\label{eq:5a}
\begin{eqnarray}
&\tau_m\dot{R}= \frac{\Delta_0J_0 }{\pi}{Y}+ 2RV  \label{rate}\\  
&\tau_m\dot{V}= V^2+ \sqrt{K}(I_0 - J_0\tau_m{Y})           -(\pi\tau_{m}R)^2\\  
& \tau_{d}\dot{{Y}} = -{Y} +R.   
\end{eqnarray}
\end{subequations}
As verified in \cite{matteo}  for instantaneous PSPs
this formulation represents a quite good guidance for
the understanding of the emergence of sustained COs in the network, despite the fact that the MF asymptotic 
solutions are always stable foci.
Instead in the present case,  analogously to what found for 
\red{structurally homogeneous networks of heterogeneous neurons in \cite{devalle2017}}, 
we observe that for IPSPs of finite duration oscillations can emerge 
in the network as well as in the mean-field, as
shown in Fig. \ref{f0}. \red{The data reported in the
figure confirm that the MF formulation \eqref{eq:5a}, despite 
not including current fluctuations, reproduces quite well the macroscopic evolution of the network
in the oscillatory regime also for a sparse network.}

Therefore we can safely employ such effective MF model 
to interpret the phenomena observed in the spiking network and to obtain theoretical predictions
for its dynamics. 

\red{
In the next two subsections we will firstly study analytically the linear stability of the asynchronous state, which corresponds to a fixed point of \eqref{eq:5a}, and then we will describe the bifurcation and phase
diagrams associated to the MF model  \eqref{eq:5a}.}

\subsection{Linear stability of the asynchronous state}

The fixed point solution $(V^*,R^*,Y^*)$ of \eqref{eq:5a} is given by:
\begin{subequations}\label{eq:6}
\begin{eqnarray}
V^{*} &=& -\frac{\Delta_{0}J_{0}}{2\pi} \enskip , \\
{R^*} \tau_m &=& \frac{J_0 \sqrt{K}}{2 \pi^2} \left(\sqrt{1+ \frac{4 \pi^2}{\sqrt{K}}\frac{I_0}{J_0^2}
+\frac{\Delta_0^2}{K}}-1\right) \enskip,\\
Y^* &=& R^* \enskip .
\end{eqnarray}
\end{subequations}
 
By performing a linear stability analysis around the fixed point solution $(V^*,R^*,Y^*)$
we obtain the following secular equation:
\begin{eqnarray}\label{eq:8}
\left| \begin{array}{ccc}
2V^{*}-\Lambda \tau_m & 2 R^* & -2 V^* \\
-2(\pi \tau_m)^{2} R^{*} & 2V^{*}-\Lambda \tau_m & -J_{0}\sqrt{K}\tau_m \\
1 & 0 & -1-\Lambda \tau_d \end{array} \right|=0.
\end{eqnarray}
in a more explicit form this is 
\begin{eqnarray}\label{eq:9}
& & \left(1+\Lambda \tau_d\right)\left[\left(\Lambda \tau_m - 2 V^*\right)^{2}+(2\pi R^{*} \tau_m)^{2}\right]
\nonumber\\
& + & 2 V^* \left( \Lambda \tau_m -2V^*\right)+ 2J_{0}\sqrt{K}R^{*}\tau_m = 0
\end{eqnarray}

In the present case, for inhibitory coupling (i.e. $J_0 >0$) the solutions of the cubic equation \eqref{eq:9}
are one real and two complex conjugates. The real one is always negative therefore irrelevant for
the stability analysis, while the couple of complex eigenvalues $\Lambda = \Lambda_R \pm i \Lambda_I$
can cross the imaginary axes giving rise to oscillatory behaviours via Hopf bifurcations. 
The presence of the two complex conjugate eigenvalues implies that whenever the asynchronous state is stable, this
is always a focus characterized by a frequency of relaxation towards the fixed point given by $\nu_{D} = \Lambda_I/2\pi$. 
For excitatory coupling, the real eigenvalue can become positive with an associated
saddle-node bifurcation and the emergence of collective chaos \cite{olmi2011,pazo2016}.

%figure 2
\begin{figure*}
\centerline{\includegraphics[scale=0.6]{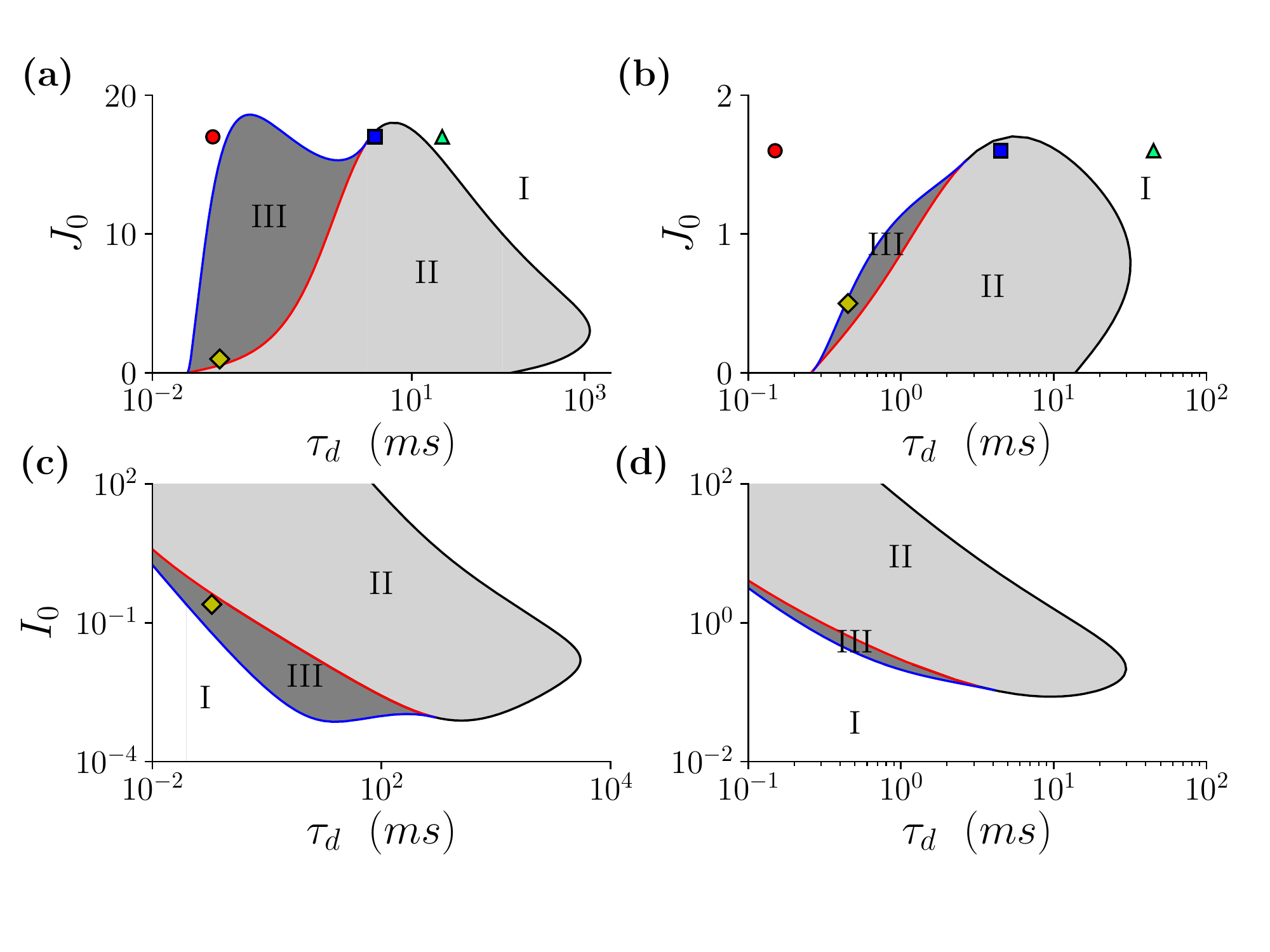}}
\caption{{\bf Phase diagrams of the mean-field model in the $(\tau_d,J_0)$-plane (a-b)
and in the $(\tau_d,I_0)$-plane (c-d)}. \red{The left panels refer to $\Delta_0 = 0.3$ 
and the right ones to $\Delta_0 = 3$. The red (black) line corresponds to sub-critical (super-critical) Hopf bifurcations, while the blue 
curve indicates saddle-node bifurcations of limit cycles.
In the region I (white) the only stable solutions are foci and in the region II (light shaded)
these are limit cycles. The dark shaded area (III) represents the region of coexistence 
of stable foci and limit cycles.The colored symbols indicate the states analyzed in Section IV. The parameters are $I_0 =0.25$ in (a-b) and
$J_0 =1.0$ in (c-d) and $K=1000$.}}
\label{f6}
\end{figure*}

By following \cite{devalle2017}, the Hopf boundaries can be identified by 
setting $\Lambda = i 2\pi \nu_O$ in \eqref{eq:9} and to zero the real and imaginary part of the resulting equation, namely one gets
 \begin{subequations}\label{eq:11}
 \begin{eqnarray}
& \frac{\left(1-4 \tau_d V^*\right)(2 \pi \nu_O)^{2}}{R^*}-(2\pi)^{2} R^* \tau_m-2 J_{0}\sqrt{K}=0\\ \label{eq:11a}
& \left[(2 \pi \nu_O)^{2}\tau_m-4(V^*)^{2}-(2\pi R^{*} \tau_m)^{2}\right] - 2 \frac{\tau_m V^*}{\tau_d} =0
\enskip . \label{eq:11b}
\end{eqnarray}
\end{subequations}

\subsection{Phase Diagrams of the Mean-Field Model}

\begin{figure*}
\centerline{\includegraphics[scale=0.6]{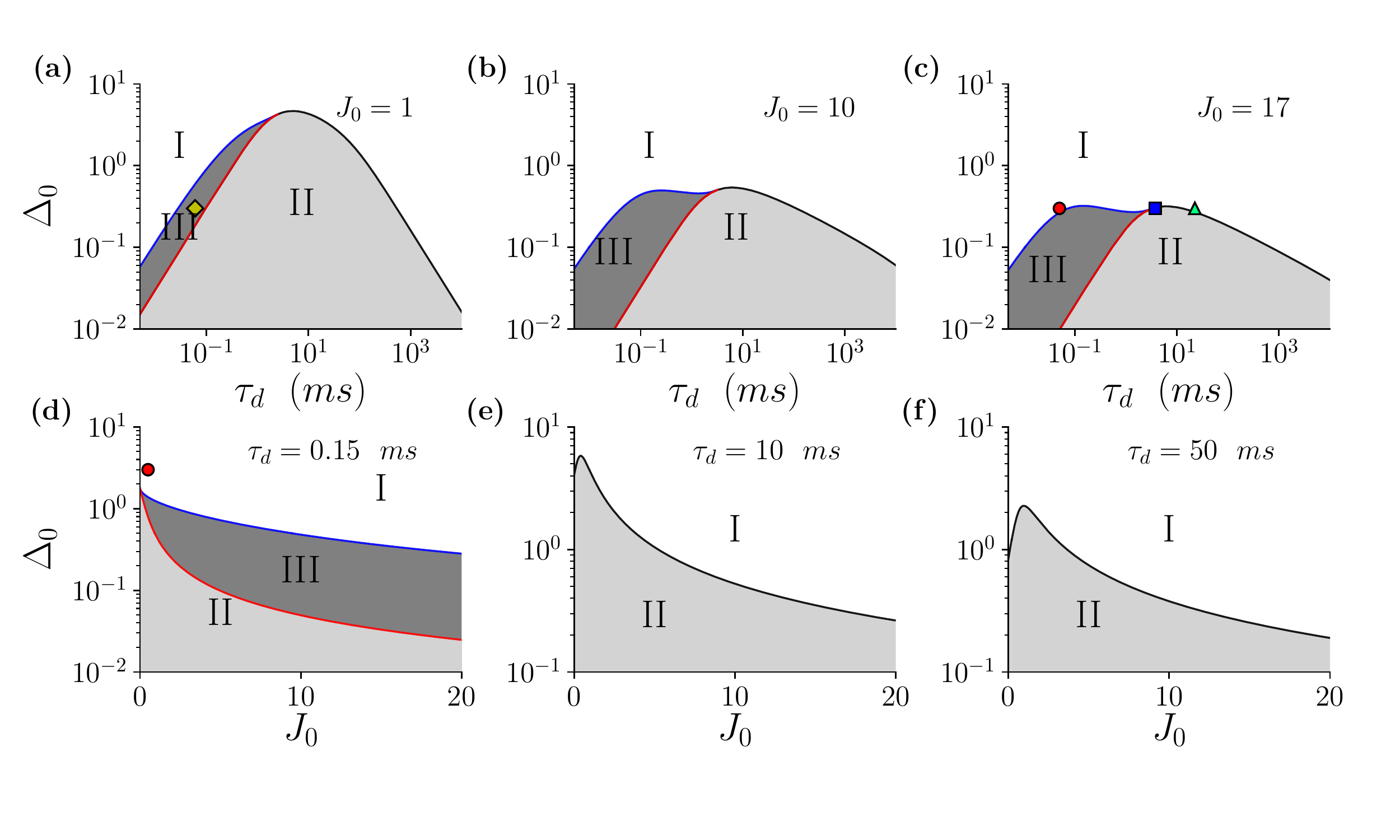}}
\caption{{\bf Phase diagrams of the mean-field model in the $(\tau_d, \Delta_0)$-plane (a-c) and
in the $(J_0, \Delta_0)$-plane (d-f)}
\red{The line colors, the colored symbols and regions are defined as in Fig. \ref{f6}.
For the parameters we fixed $I_0 = 0.25$ and K=1000.
}}
\label{f7}
\end{figure*}

\red{
Apart from the linear stability of the asynchronous state and the
associated Hopf boundaries which can be worked out
analytically, the limit cycle solutions of the MF
model and the associated bifurcations have been
obtained by employing the software XPP AUTO 
developed for orbit continuation \cite{XPP2007}. 
The MF model  \eqref{eq:5a}, apart from
the membrane time constant $\tau_m$, which sets the 
system time scale, and the median in-degree $K$, which we
fixed to 1000, is controlled
by four independent parameters: namely, $\Delta_0$, $J_0$, $I_0$, $\tau_d$.
In the following we will give an overview of the possible behaviours 
of the MF model in terms of two parameters phase diagrams for the most relevant 
combinations of the four mentioned parameters. The results
of these analysis are summarized in Figs. \ref{f6} and \ref{f7}. 
}

\red{Our analysis of the stationary solutions has revealed three possible regimes:
stable foci (I); stable COs (II); coexistence of these two stable solutions (III).
The stability boundaries of the COs are delimited by three kind of bifurcations:
super-critical Hopf (black lines in the figures); sub-critical Hopf (red lines) 
and saddle-node (SN) of limit cycles (blue lines). Stable (unstable) COs emerge from stable
foci at super-critical (sub-critical) Hopfs, while stable and unstable limit cycles
merge at the SNs. 
}

\red{A fundamental parameter controlling the emergence of COs in the MF model
is the synaptic time $\tau_d$, indeed in absence of this time scale no oscillations are
present at the MF level \cite{matteo}. On the other hand too large values of $\tau_d$ 
also lead to COs suppression, since the present model reduces to a Wilson-Cowan model
for a single inhibitory population, that it is know to be unable to display oscillations \cite{devalle2017}.
As shown in Figs. \ref{f6} and \ref{f7}, oscillations are observable for intermediate values of $\tau_d$
and not too large $J_0$, since large inhibition leads to a quite reduced activity
of the neurons not sufficient to ignite a collective behaviour.
This is in agreement with the fundamental role played by gamma-Aminobutyric acid (GABA)
in the emergence of epileptic seizures, characterized by an anomalous level
of synchronization among the neurons, indeed the occurence of seizures
seems strongly correlated with a GABA deficit, corresponding to a 
reduction of $J_0$ in our case \cite{sperk2004,gonzalez2015}.
Moreover, in order to observe
COs the excitatory drive $I_0$ should be larger than some critical 
value, as shown in Fig. \ref{f6} (c-d).
This is consistent with the observation of the emergence of gamma oscillations in hippocampal
slices induced through the acetylcoline agonist charbachol \cite{fisahn1998,dickinson2003},
which leads to a decrease of the conductances of potassium channels, which can be mimicked
as an increase of $I_0$ \cite{mccormick1985,frazier1998}. Indeed, by increasing the structural 
heterogeneity (measured by $\Delta_0$),
which acts against coherent dynamics, larger values of $I_0$ are
required for COs as well as smaller synaptic couplings (see Figs. \ref{f6} (b),(d)
and \ref{f7} (d-f)).
Therefore the emergence of COs can be triggered by self-disinhibition as well as by an external excitatory drive,
and we expect to observe in both cases the same scenarios. 
}

\red{ 
As already mentioned, for infinitely fast synapses ($\tau_d \to 0$) the
only possible solutions of the MF are foci characterized by two complex
conjugate eigenvalues. Nevertheless, in the corresponding network
the irregular firings of the neurons, due to the
dynamical balance, can sustain COs, which are predicted to relaxed toward the fixed point in the MF. 
In the next Section we will analyze
the role of these microscopic fluctuations in triggering the network dynamics
also for finite $\tau_d$.}

%figure 3

\section{Network dynamics}

We investigate in this Section the dynamics of the network by considering the parameter plan $(\tau_d,J_0)$. 
In particular, we want to examine the role of structural heterogeneity (measured by $\Delta_0$) in shaping the dynamical behaviours. This characteristic of the network structure is extremely relevant, as it can determine even if the system is in a balanced or in an imbalanced regime \cite{pyle2016, landau2016,matteo}.

\subsection{High structural heterogeneity}

\begin{figure}
\centerline{\includegraphics[scale=0.45]{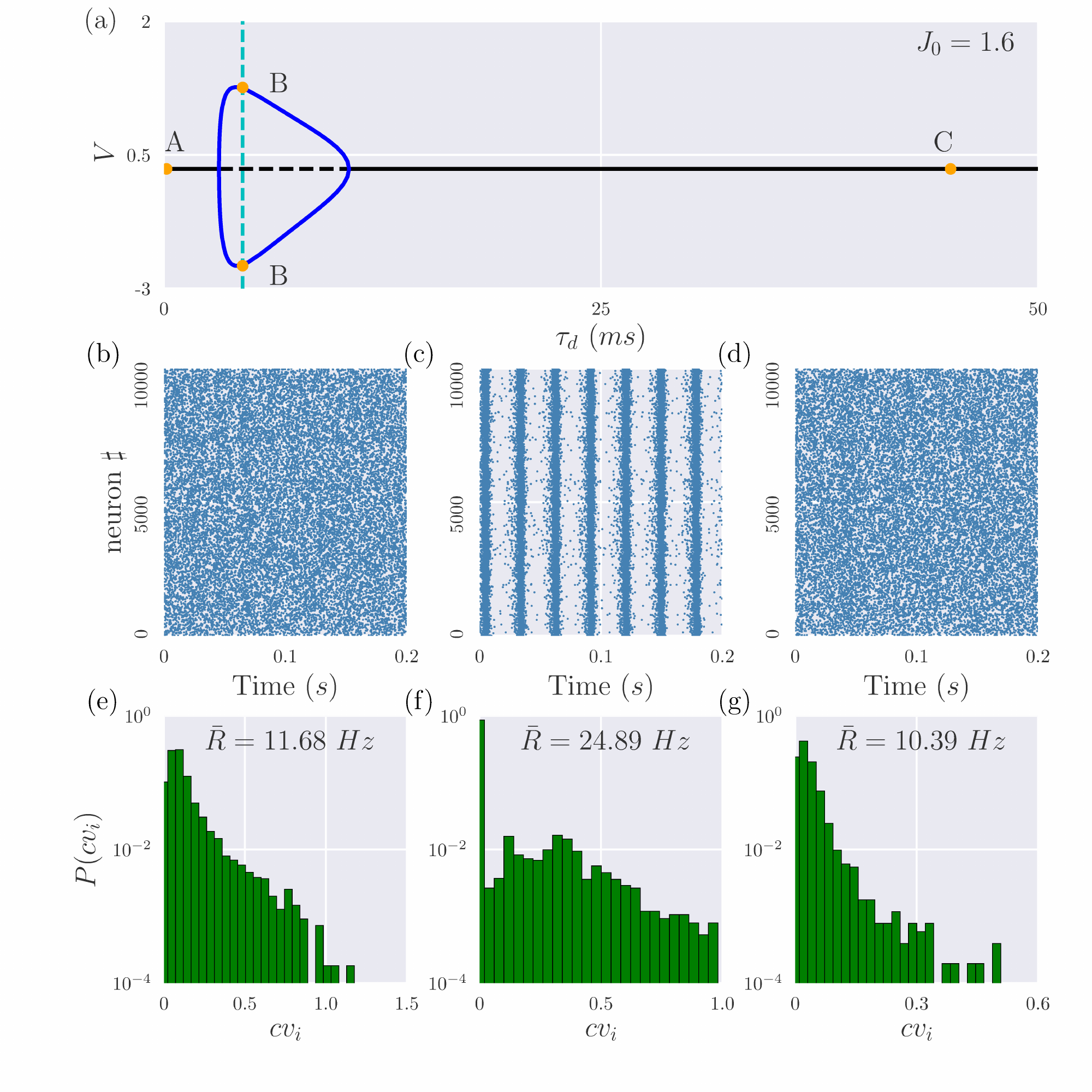}}
\caption{{\bf High Structural heterogeneity: super-critical Hopf bifurcation.}
(a) Bifurcation diagram of the MF model \eqref{eq:5a}
displaying the extrema of $V$ versus $\tau_d$,
black solid (dashed) lines refer to the stable (unstable) focus,
while blues solid lines to the oscillatory state.
 \red{The supercritical Hopf bifurcations take place for 
$\tau^{(H)}_1 =3.14$ ms and $\tau^{(H)}_2 = 10.59$ ms.}
The capital letters in (a) denote three stationary states corresponding
to different synaptic time scales, namely: (A) $\tau_d=0.15$ ms;
(B)  $\tau_d=4.5$ ms and (C) $\tau_d=45$ ms. The network dynamics
corresponding to these states is reported in the panels below:
the left column corresponds to (A),  the central to (B) and the
right one to (C). For each column, the top panels are the corresponding
raster plots (b,c,d) and the bottom ones the distributions of the $\{cv_i\}$ of the single neurons
(e,f,g). Network parameters are $N = 10000$, $K = 1000$ and $\Delta_0 = 3.0$.  Other parameters are $I_0 = 0.25$ and $J_0 = 1.6$. 
}
\label{f1}
\end{figure}

We consider first a relatively high value for the structural heterogeneity,
namely $\Delta_0=3.0$. 
For sufficiently large coupling $J_0$, the bifurcation diagram reveals the emergence of oscillations in the MF model (\ref{eq:5a}) via super-critical Hopf bifurcations, analogously to what has been reported for globally coupled networks  \cite{devalle2017}.
An example of the bifurcation diagram, displaying the extrema of the mean membrane potential $V$ as a function of $\tau_d$ is reported in Fig. \ref{f1} (a) for $J_0=1.6$. 
In particular, we observe for instantaneous synapses ($\tau_d \to 0$) a stable focus, as expected from the analysis previously reported in \cite{matteo}. 
The focus is stable up to $\tau^{(H)}_1$ where it is
substituted by a stable oscillatory state via a super-critical Hopf
bifurcation. Oscillations are observable up to $\tau^{(H)}_2$,
where via a second super-critical Hopf bifurcation they disappear and
the unique stable solution for the MF system remains a focus.
The typical stable regimes are denoted in Fig. \ref{f1} (a) by three capital
letters: namely, (A) corresponds to a focus,
(B) to a limit cycle and (C) to another focus. The network dynamics corresponding to these typical MF
solutions is examined in the remaining panels of Fig. \ref{f1}.
For the focus solutions the network dynamics is asynchronous,
as clearly visible from the corresponding raster plots in Fig. \ref{f1} (b) and (d).
Furthermore, the dynamics of the neurons is quite regular in this case,
as testified from the values of the average coefficients of variation, namely ${CV} \simeq 0.14$ and ${CV} \simeq 0.04$ corresponding to the distributions reported in Fig. \ref{f1} (e) and (f), respectively. 
At intermediate values of $\tau_d$, as predicted by the MF analysis, we observe COs 
with frequency $\nu_{OSC} \simeq 34$ Hz in the network dynamics, see Fig. \ref{f1} (c). However, also in this case the dynamics is dominated by supra-threshold neurons
with an associated very low $CV$, as evident from the large peak present at $cv_i \simeq 0$ 
in the distribution $P(cv_i)$ shown in Fig. \ref{f1} (g).

%figure 4

\begin{figure}
\centerline{\includegraphics[scale=0.35]{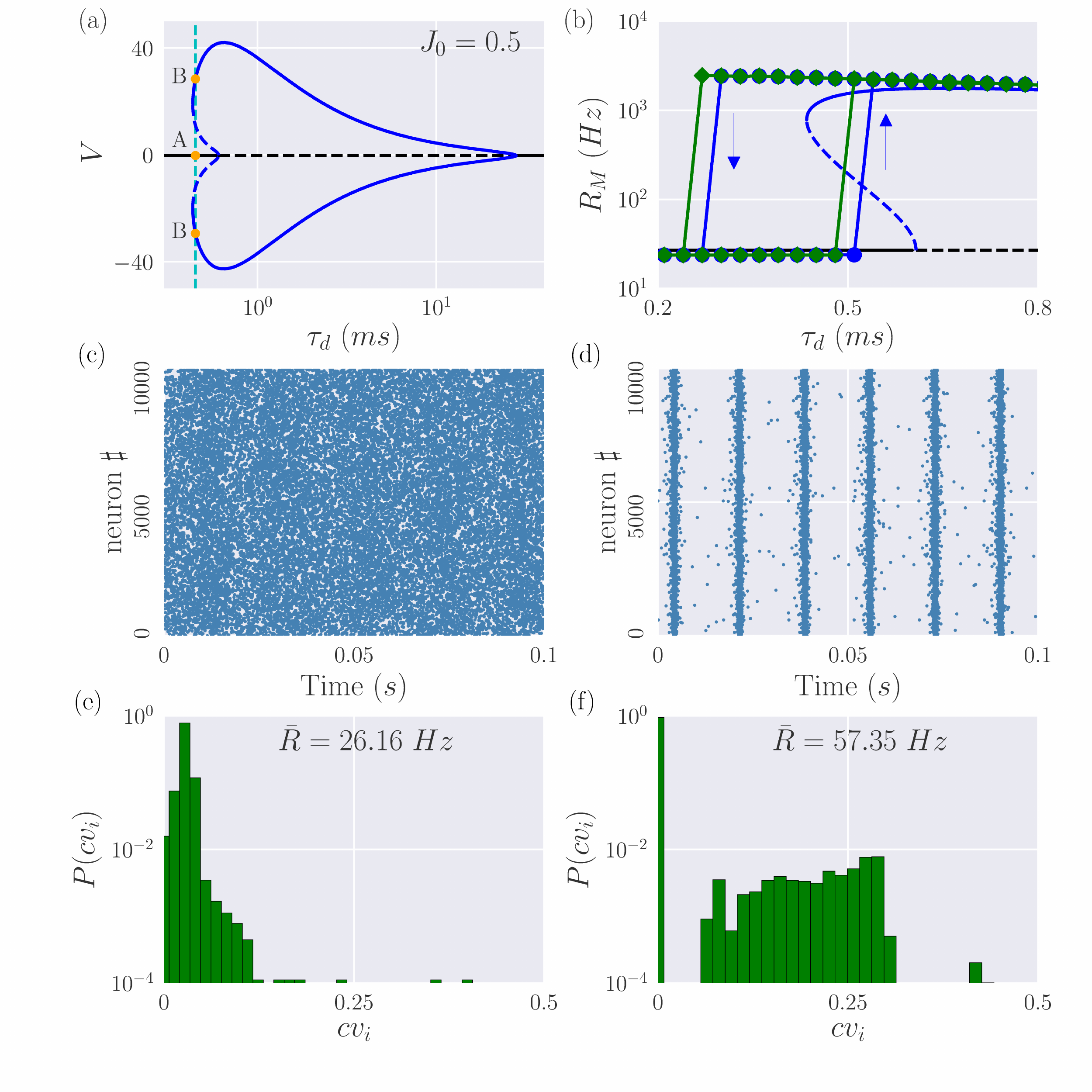}}
\caption{{\bf High structural heterogeneity: sub-critical Hopf bifurcation.} 
(a) Bifurcation diagram of the MF model analogous to the one reported in
Fig. \ref{f1} (a). \red{The super-critical (sub-critical) Hopf bifurcation takes place at 
$\tau^{(H)}_2 = 27.96$ ms ($\tau^{(H)}_1 = 0.61$ ms), the saddle node
of limit cycles at $\tau^{(S)} = 0.43$ ms.}
The capital letters in (a) denote two stationary states corresponding
to the same synaptic time scales $\tau_d=0.45$ ms.
The network dynamics corresponding to these states is reported in the panels below:
the left column correspond to (A) and the
right one to (B). For each column, the top panels display the
raster plots (c,d) and the bottom ones the distribution of the $\{cv_i\}$ of the single neurons (e,f).
In panel (b) are reported the maximal values of the rate $R_M$ obtained by performing adiabatic simulations by first increasing and then decreasing the synaptic time $\tau_d$ (green) diamonds for $N=10,000$ and (blue) circles for $N=20,000$, the arrows denote the jump from one state to the other.
The MF results are also displayed: solid (dashed) black lines refer
to stable (unstable) foci, while solid (dashed) blue lines
to stable (unstable) limit cycles.
Parameters are the same as in Fig. \ref{f1}, apart for $J_0 = 0.5$,
\red{the parameters for the adiabatic simulations are 
$\Delta \tau_d = 0.03$ ms, $\tau_d^{(0)} = 0.21 $ ms and $\tau_d^{(1)} = 0.81$ ms.}
}
\label{f2}
\end{figure}

For lower synaptic coupling $J_0$ the phase portrait changes,
as shown in Fig. \ref{f2} (a) for $J_0=0.5$. In this case the MF analysis
indicates that the transition from a stable focus to the oscillatory
state occurs by increasing $\tau_d$ via a sub-critical Hopf bifurcation.
At large synaptic coupling, the stable focus is recovered via a super-critical Hopf bifurcation taking place at $\tau^{(H)}_2$, analogously to what has been seen for larger coupling. An interesting regime is
observable between $\tau^{(S)}$, where the stable and unstable limit cycle merge via a saddle-node bifurcation, and $\tau^{(H)}_1$, where the focus become unstable. In this interval the MF model displays
two coexisting stable solutions: a limit cycle and a focus.
It is important to verify if also the finite size
sparse network displays this coexistence, indeed as shown in Fig. \ref{f2}
depending on the initial conditions the network dynamics 
can converge towards COs or towards an asynchronous state. In particular, we
observe that the asynchronous dynamics is associated to extremely low $cv$-values (see Fig. \ref{f2} (d)) suggesting that this can be considered as a sort of irregular splay state \cite{olmi2012stability}. However,
also the COs with $\nu_{OSC} \simeq 58$ Hz are characterized by a low average coefficient of variation,
namely ${CV} \simeq 0.014$ indicating that the dynamics is mean driven.
The sub-critical Hopf, as expected, is associated to a histeretic behaviour,
this effect can be revealed by considering simulations concerning an adiabatic variation of $\tau_d$.
The results of these simulations are reported in Fig.  \ref{f2} (b), where the maximal values
of the instantaneous firing rate $R_M$ are reported as a function of $\tau_d$ for the adiabatic protocol and compared with
the MF estimations of $R_M$.  From the figure it is clear that the transition from the focus to the
stable limit cycle occurs at $\tau_d < \tau^{(H)}_1$ and the system returns from the oscillatory state to the asynchronous one at $\tau_d$ definitely smaller than $\tau^{(S)}$. These are finite size (and possibly also finite time) effects, indeed as shown in Fig.  \ref{f2} (b) by increasing $N$ the transition points approach the MF ones.

%figure 5

\subsection{Low structural heterogeneity}

\begin{figure}
\centerline{\includegraphics[scale=0.45]{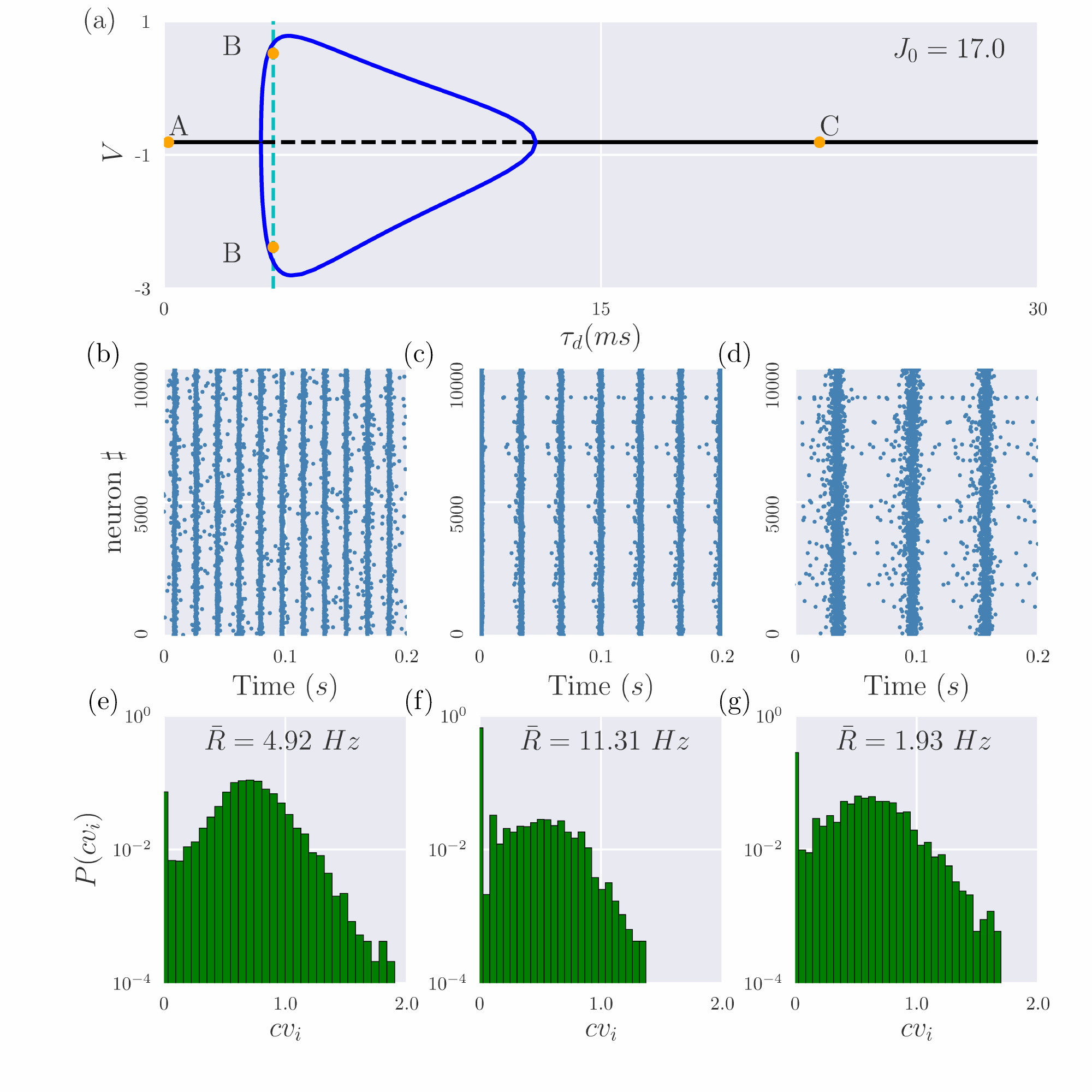}}
\caption{{\bf Low structural heterogeneity: super-critical Hopf bifurcation.} 
The panels here displayed are  analogous to the ones in Fig. \ref{f1}.
In this case \red{the super-critical Hopf bifurcations
occur for $\tau^{(H)}_1 = 3.33$ ms and $\tau^{(H)}_2 =  12.61$ ms and}
the stationary states in (a) corresponding to the capital letter
(A), (B) and (C) refer to  $\tau_d=0.15$ ms, $\tau_d=3.75$ ms and $\tau_d=22.5$ ms,
respectively. The parameters are the same as in Fig. \ref{f1}, apart
$\Delta_0 = 0.3$.  and $J_0 = 17$. 
}
\label{f3}
\end{figure}

We consider now a relatively low value of the structural heterogeneity, i.e. $\Delta_0=0.3$, 
which for instantaneous synapses can sustain a dynamically balanced state~\cite{matteo}.
Let us first consider a relatively large coupling,
namely $J_0 = 17.0$, the corresponding bifurcation diagram
for the MF model is reported in Fig. \ref{f3}(a). 
This is quite similar to the one previously shown for 
high structural heterogeneity in Fig. \ref{f1}(a).
However, peculiar differences are observable at the level of network simulations.
Indeed in this case COs are present for all the considered $\tau_d$-values, even
if these correspond to stable foci in the MF (states (A) and (C) in Fig. \ref{f3}(a))
as evident from the raster plots reported in Fig. \ref{f3}(b) and (d).
In particular we measured the following frequencies for the observed COs:
$\nu_{OSC} \simeq 57$ Hz for state (A), $\nu_{OSC} \simeq 30$ Hz for (B)
and $\nu_{OSC} \simeq 16$ Hz for (C). Furthermore, the network dynamics is now definitely more irregular
than for high $\Delta_0$ with distributions $P(cv_i)$ 
centered around $cv_i = 1$ for the states (A) and (C) in Fig. \ref{f3}(a) corresponding
to stable foci in the MF formulation (see Fig. \ref{f3}(e) and (g))
and with $P(cv_i)$ extending towards values around $cv_i \simeq 1$  for the
oscillatory state (B), as shown in Fig. \ref{f3}(i). 
This irregularity in the spike emissions is a clear indication that
now the dynamics is mostly fluctuation driven due to the dynamically balanced dynamics
observable in the sparse network for sufficiently low structural heterogeneity. 
Furthermore, as shown in \cite{matteo} for instantaneous synapses,
these current fluctuations are able to turn the macroscopic damped oscillations towards the stable foci,
observable in the MF model, in sustained COs in the network.
The origin of the COs observable for the state (B) is indeed different, 
since in this case sustained oscillations emerge due to a super-critical Hopf bifurcation
both in the MF and in the network dynamics.

%figure 6

\begin{figure}
\centerline{\includegraphics[scale=0.35]{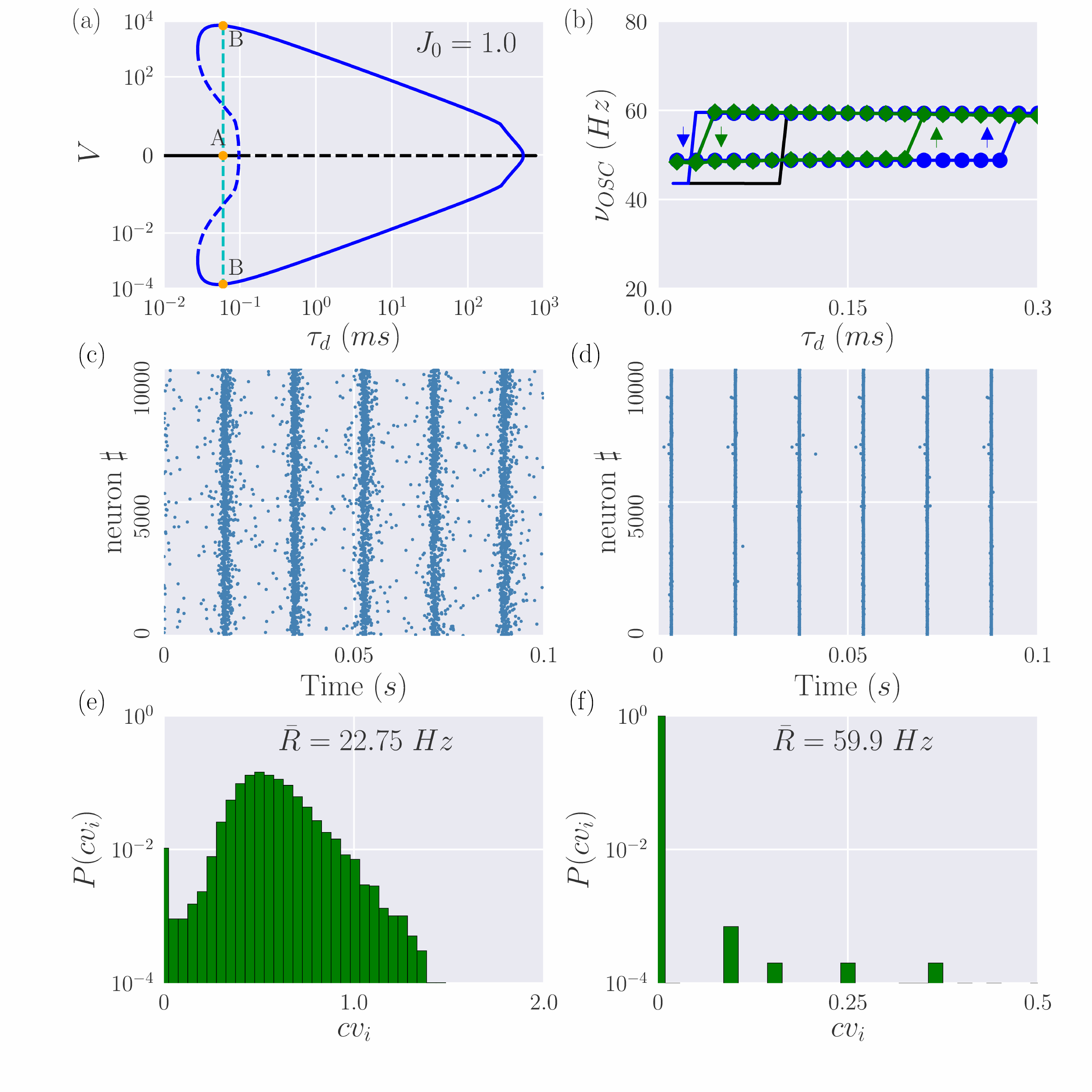}}
\caption{{\bf Low structural heterogeneity: sub-critical Hopf bifurcation.}
The panels here displayed, apart panel (b), are  analogous to the ones in Fig. \ref{f2}.
In the present case \red{the sub-critical Hopf occurs at
$\tau^{(H)}_1 = 0.097$ ms, while the super-critical Hopf at $\tau^{(H)}_2 = 531.83$ ms
and the saddle-node of limit cycles at $\tau^{(S)} = 0.028$ ms, and }
the coexisting states (A) and (B) shown in (a)  
refer to  $\tau_d = 0.06$ ms. Panel (b)  reports the frequency
of collective oscillations as measured via adiabatic simulations for $N=2000$
by considering $T_s = 90$ ms (blue circles) and $T_s = 1500$ ms (green diamonds),
the transient time $T_t = 15$ ms is unchanged. The  solid lines in (b) refer to the MF results,
namely the black line to $\nu_{D}$ and the blue one to the limit cycle frequency $\nu_O$.
The parameters are as in Fig. \ref{f3} apart for $J=1.0$ \red{and for the
adiabatic simulations are $\Delta \tau_d = 0.015$ ms, $\tau_d^{(0)} = 0.015 $ ms and $\tau_d^{(1)} = 0.30$ ms.}
}
\label{f4}
\end{figure}

By decreasing the synaptic coupling $J_0$ (Fig. \ref{f4}(a)) we observe in the MF
phase diagram the emergence of regions where the oscillations coexist with the stable focus in proximity of a sub-critical Hopf bifurcation, analogously to what has been reported for high heterogeneity (see Fig  \ref{f2}(a)).  
At variance with that case, we have now in the network a bistability 
between two COs whose origin is different: one emerges via a Hopf bifurcation
and it is displayed in Fig.  \ref{f2}(d), while
the other is sustained by the irregular spiking associated to the balanced state
and the corresponding raster plot is reported in Fig.  \ref{f2}(c).
In particular, the latter
COs are associated to large $cv$-values (Fig \ref{f4} (e)) typical of
a balanced regime, while the other COs are extremely regular as shown in
Fig \ref{f4} (f) resembling the dynamics of a highly synchronized system.

%figure 7

In order to analyze the coexistence region, we report 
in Fig. \ref{f4}(b) the frequencies $\nu_{OSC}$ of the collective oscillations 
as measured via adiabatic simulations of the network (symbols).
Furthermore, the MF results for $\nu_{D}$ associated to the foci
and the frequencies $\nu_O$ of the limit cycles are also reported in the figure as black and blue solid
line, respectively. The frequencies of the COs in both states are
reasonably well captured by the MF approach, furthermore the two frequencies 
can be quantitatively associated to fast and slow gamma oscillations.
The comparison reveals that the 
COs induced by microscopic irregular firing exist far beyond 
$\tau^{(H)}_1$, despite here the unique stable solution predicted by the MF should be 
the almost synchronized bursting state shown in Fig. \ref{f4}(d).
On the other hand the backward transition is almost coincident with the
MF prediction for $\tau^{(S)}$ as displayed in Fig. \ref{f4}(b).
As reported in Fig. \ref{f4}(b),
we observe that also the forward transition value approaches to $\tau^{(H)}_1$
by increasing the duration $T_s$ of the adiabatic steps.
Therefore this result suggest that the observed discrepancies are due to finite time
(and possibly finite size) effects affecting the network simulations.

\section{Coexistence of slow and fast gamma oscillations}

\begin{figure}
\centerline{\includegraphics[scale=0.45]{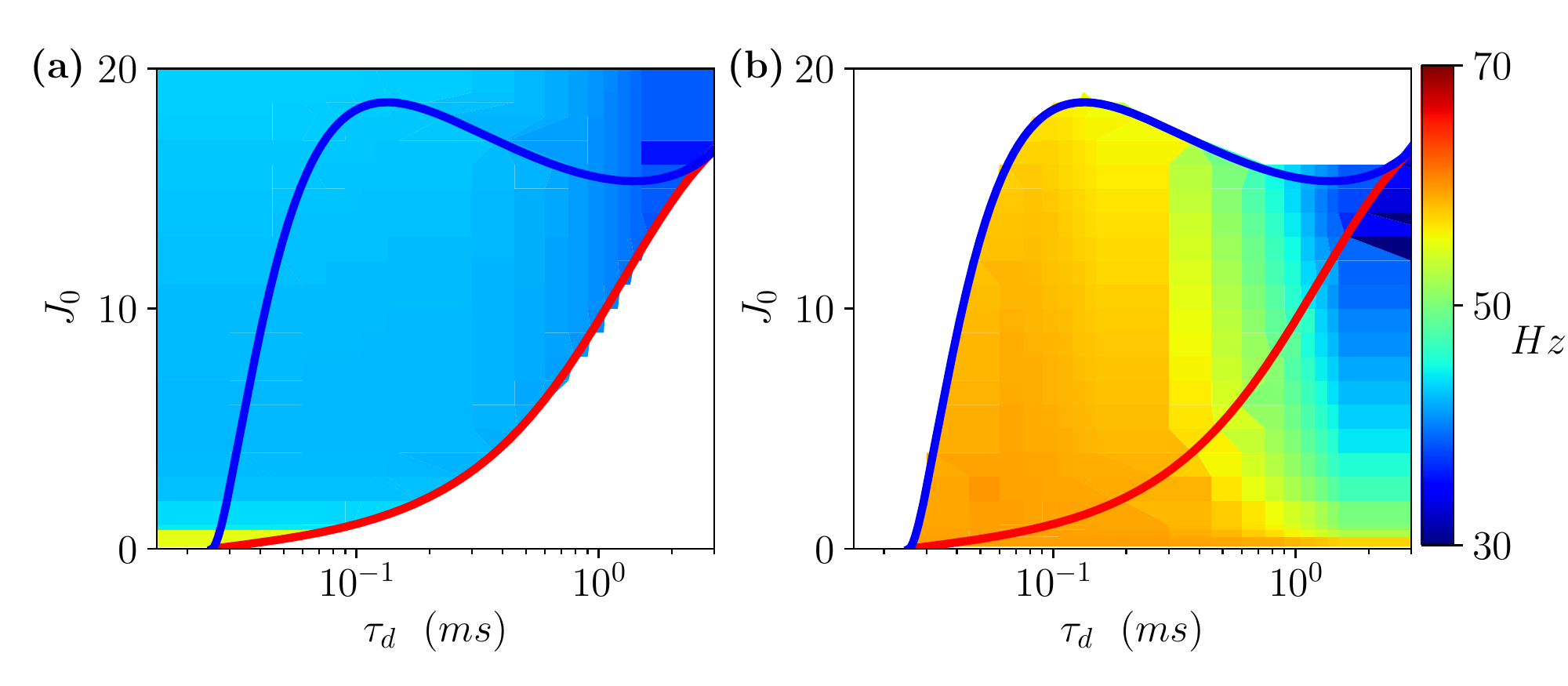}}
\caption{{\bf Coexisting fast and slow gamma oscillations} 
(a) Frequencies $\nu_D$ associated to the damped oscillations towards the stable foci;
(b) frequencies  $\nu_O$ of the limit cycles.
Red lines refer to the sub-critical Hopf boundaries, while
the blue ones to saddle-node bifurcations of limit cycles.
Parameters as in Fig \ref{f6}. }
\label{f9}
\end{figure}

In the previous Section we have shown, for a specific choice of the parameters,
that fast and slow collective gamma oscillations can coexist.
However, the phenomenon is observable in the whole range
of coexistence of the stable foci and of the stable limit cycles.
In particular, in Fig. \ref{f9} we report in the $(\tau_d,J_0)$-plane
the frequencies $\nu_D$ associated to the damped oscillations towards the MF focus 
in panel (a) and the frequencies $\nu_O$ of the 
limit cycles in panel (b). It is evident that $\nu_{D} \simeq 30-40$ Hz,
while the frequencies of the limit cycle $\nu_O$ are of the order of $60$ Hz,
thus in the network we expect to observe coexisting COs characterized by
slow and fast  rhythms in a wide range of parameters.

\red{For this parameter set $\nu_D$ seems to depend only slightly
on $\tau_d$ and $J_0$. On the contrary the frequency $\nu_O$,
characterizing the more synchronized events, is influenced by these parameters.
In particular, $\nu_O$ decreases for increasing IPSP  time duration, 
analogously to what observed experimentally for cholinergic induced gamma oscillations in 
the hippocampus  {\it in vitro} \cite{fisahn1998}. 
Moreover, barbiturate, a drug often used as anxiolytic, is konwn to increase IPSP
time duration \cite{nicoll1975} and slow down gamma oscillations \cite{traub1998}, in accordance
with our scenario. Furthermore, for $\tau_d > 1$ ms the increase of $J_0$ leads to a decrease of $\nu_D$,
similarly to the effect of alcohol that induces an increase of inhibition
associated to a decrease in gamma oscillation frequencies measured
in the human visual cortex \cite{campbell2014}.
}

\red{The coexistence of fast and slow gamma COs is a quite general phenomenon
not limited  to the specific network topology we employed, i.e. that associate to the Lorentzian 
in-degree distribution. Indeed, as shown in Appendix A it can be observed also
for a sparse Erd\"os-Renyi network.}

\subsection{Switching gamma rhythms}

\begin{figure}
\centerline{\includegraphics[scale=0.45]{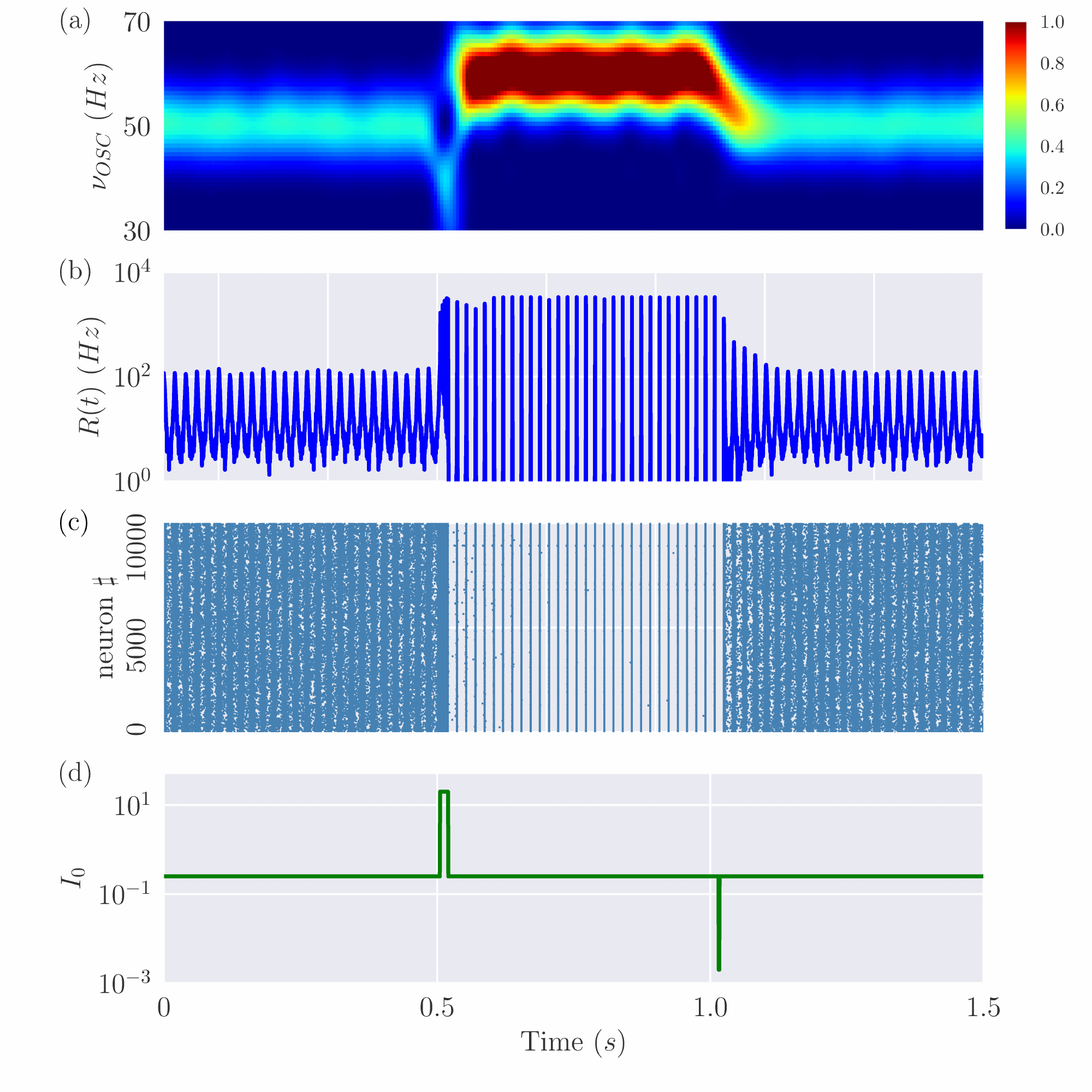}}
\caption{{\bf Switching from fast (slow) to slow (fast) gamma oscillations}
Results of the switching experiments described in the text ,
from top to bottom: (a) spectrogram of the mean membrane potential $V$;
(b) the firing rate $R(t)$; (c) the raster plot and (d) the stimulation protocol
reporting the average external DC current.
The parameters are the same as in Fig. \ref{f4} (in particular $\tau_d = 0.06$ ms), 
apart $T_{sh}=0.015$ s, $T_{sl}=0.0015$ s, $I_{1}=0.25$, $I_{2}=20.0$, $I_{3}=0.012$,
$\Delta I_3 = 0.01$. }
\label{f11}
\end{figure}

As a further aspect, we will consider the possibility to develop a simple protocol
to drive the system from slow gamma COs to fast ones (and vice versa) in the bistable regime. Let us consider the case where the network is oscillating
with slow gamma frequency as shown in Fig. \ref{f11} for $I_0 \equiv I_1 = 0.25$. The protocol to drive the system in the fast gamma band consists in delivering a step current $I_2$ to all the neurons
for a very limited time interval $T_{sh}$. In this way the system is transiently driven in a regime where oscillatory dynamics is the only stable solution, as a matter of fact the neurons remain in a
high frequency state even after the removal of the stimulation, when $I_0$ returns to the initial value $I_1$ (see Fig. \ref{f11}). In order to desynchronize the neurons and to recover the slow gamma COs , we delivered
random quenched DC currents 
$I_0(i)$ (with $i=1,\dots,N$) to the neurons for a time period $T_{sl}$. The currents $I_0(i)$ are taken from a flat distribution 
with a very low average value $I_3$ and a width $\Delta I_3$ corresponding to a parameter range
where the MF foci are the only stable solutions. As shown in Fig. \ref{f11}
in this case to drive the system from fast to slow gamma oscillations it was sufficient to apply the perturbation
for a much smaller period $T_{sl} << T_{sh}$.

%figure 9

\begin{figure}
\centerline{\includegraphics[scale=0.45]{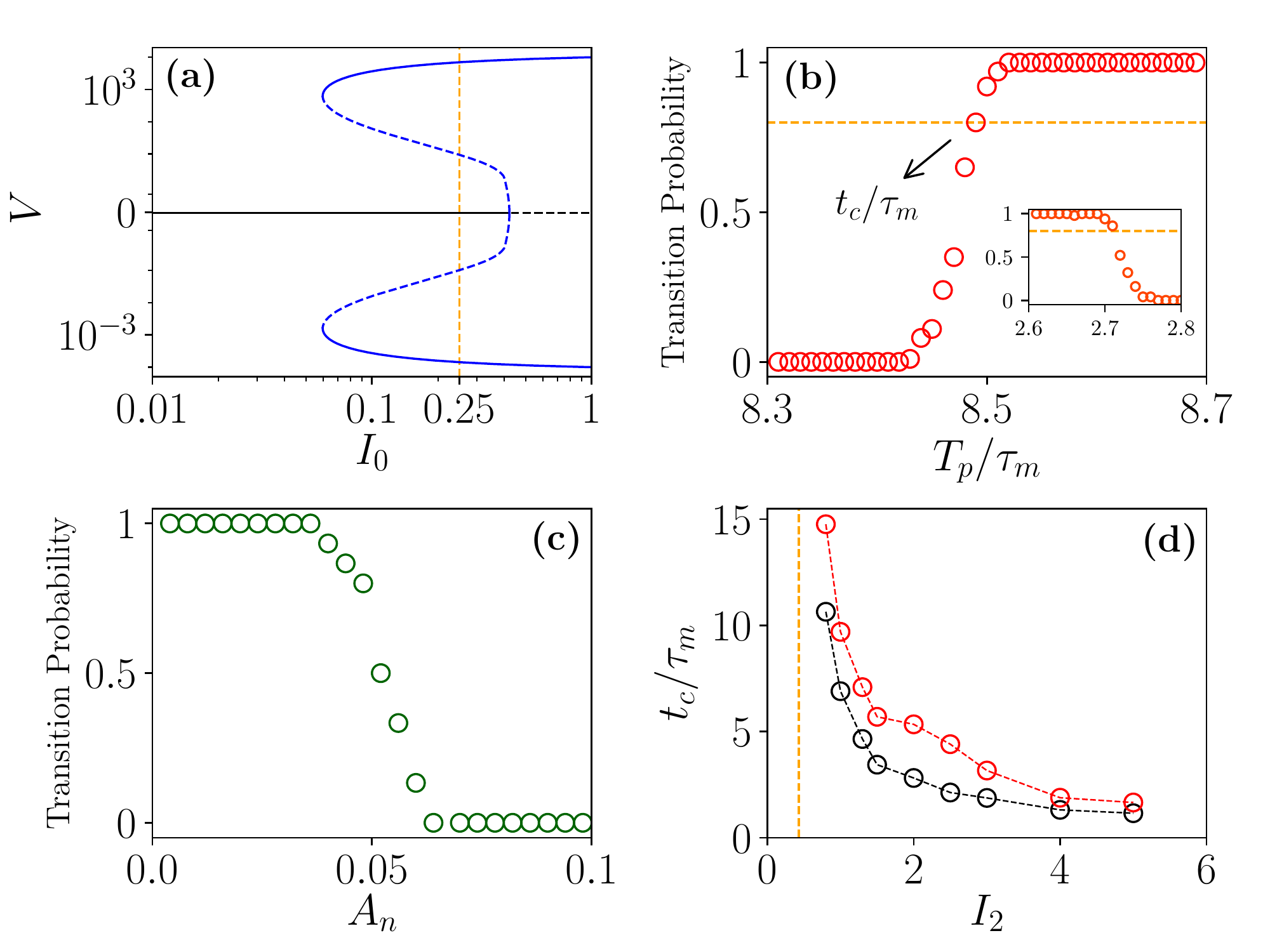}}
\caption{{\bf Statistics of the switching transitions}
\red{
(a) Bifurcation diagram of the MF model reporting the extrema of the mean membrane potential 
$V$ as a function of $I_0$ displaying stable (solid line) and unstable solutions (dashed lines) 
for foci (black) and limit cycles (blue). The vertical dashed (orange) line
refers to $I_0 = 0.25$. (b) Transition probability 
as a function of $T_P$, the orange dashed line denotes the 80 \% for $I_2=1.0$,
in the inset the data for the transition from fast to slow gamma is reported for $I_3=0.03$, in both
cases $A_n=0.05$ (c).
Transition probability as a function of the noise amplitude $A_n$ for $I_2 =1$ and $T_P=8.48 \tau_m$.
(d) Crossing times $t_c$ versus the perturbation amplitude $I_2$ for various
noise levels: $A_n= 0.02$ (black circles) and 0.07 (red circles). The vertical
orange line indicates the value $I^{(H)}$.
Panels (c-d) refer to the transition from slow to fast gamma COs, while the
inset in (b) to the transition from fast to slow gamma.
The parameters are the same as in Fig. \ref{f4}.}}
\label{f15}
\end{figure}

\red{
Let us now try to characterize in more details the observed
switching transitions.  This can be done by considering the MF bifurcation 
diagram in terms of the external DC current $I_0$
reported in Fig. \ref{f15} (a) for the examined parameters.
The diagram reveals a sub-critical Hopf bifurcaton taking place at
$I^{(H)} \simeq 0.43$ and a region of bistability extending
from $I^{(S)} \simeq 0.06$ to $I^{(H)}$.
Therefore, if we consider a DC current in the bistable interval (namely, $I_0 \equiv I_1 = 0.25$)
and we prepare the system in the slow gamma regime a transition to the
fast gamma COs will be observable whenever the DC current is increased
to a value $I_0 \equiv I_2 > I^{(H)}$. However, if we return in the bistable
regime at $I_0 \equiv I_1$, after delivering the perturbation $I_2$ for a time interval $T_P$,
it is not evident in which regime (fast or slow) the system will end up.
Thus we have measured the transition probability from slow to 
fast gamma for different $T_P$ and $I_2$ by following the protocol reported in Section II B.
We analized these transitions in presence of a small additive noise
on the membrane potentials of amplitude $A_n$, somehow encompassing the
many sources of noise present in neural circuits.
}

\red{
The results shown in Fig. \ref{f15} (b) for $I_2=1.0$ and $A_n = 0.05$
reveal that even if $I_2 > I^{(H)}$ the perturbation should be applied 
for a minimal time interval $T_P > t_c \simeq 0.12$ s to induce the transition
to the fast gamma COs in at least the 80\% of cases. It is interesting to
note that the noise amplitude can play a critical role on the switching transition,
indeed the increase of $A_n$ can desynchronize the fast gamma regime 
even for $T_P > t_c$, see Fig. \ref{f15} (c). Therefore $t_c$ depends critically 
not only on $I_2$ but also on $A_n$: as expected by increasing $I_2$
the crossing time drops rapidly towards zero, while the switching transition
is delayed to longer times for larger $A_n$ (see Fig.  \ref{f15} (d)).
}

\red{For what concerns the transition from fast to slow gamma, this occurs
in an irreversible manner only for amplitude of the perturbation $I_3 < I^{(S)} $, 
an example is reported in the inset of Fig. \ref{f15} (b).  Despite the switching 
transition can be observed also for   $I_3 > I^{(S)} $
this will be much more complex due to the competitionon of the two stable states
in the interval $I_0 \in [I^{(S)} : I^{(H)}] $ and more specific protocols should be
designed to obtain the desynchronization of the system.}

%figure 9

\subsection{Theta-gamma cross-frequency coupling}

So far we have described a simple protocol where external constant stimulations to the inhibitory
network can drive the neural population from one state to the other. \red{However,
gamma oscillations are usually modulated by theta oscillations in the hippocampus and in the neocortex
during movement and REM sleep \cite{buzsaki2002,sirota2008}. This has recently inspired a series of
optogenetic experiments {\it in vitro} intended to reproduce the effect
of the theta forcing and the activity observed {\it in vivo} \cite{akam2012,pastoll2013,butler2016}.
To make a closer contact with these experiments we decided to consider a periodic
stimulation of all neurons in the network as follows:}
\begin{equation}
I_0 (t) = I_\theta [1 - cos(2 \pi \nu_\theta t)] \quad ;
\label{forcing}
\end{equation}
where the phase of the theta forcing is defined as $\theta(t) = 2 \pi \nu_\theta t$.
The term appearing in \eqref{forcing} corresponds to the synaptic input
received by the neurons, in order to compare this forcing term with the
LFPs experimentally measured in \cite{colgin2009,belluscio2012}
and which reveals theta oscillations, one should remember that the LFP
corresponds to the electrical potential measured in the extracellular 
medium around neurons \cite{nunez2006}. Therefore for a meaningful comparison with the
synaptic input \eqref{forcing} the sign of the LFP should
be reversed. This is consistent with the observations reported 
in \cite{colgin2009,belluscio2012} that the maximum of activity of the excitatory (pyramidal) 
cells is observed in correspondence of the minimum of the LFP.

\begin{figure*}
\centerline{\includegraphics[scale=0.6]{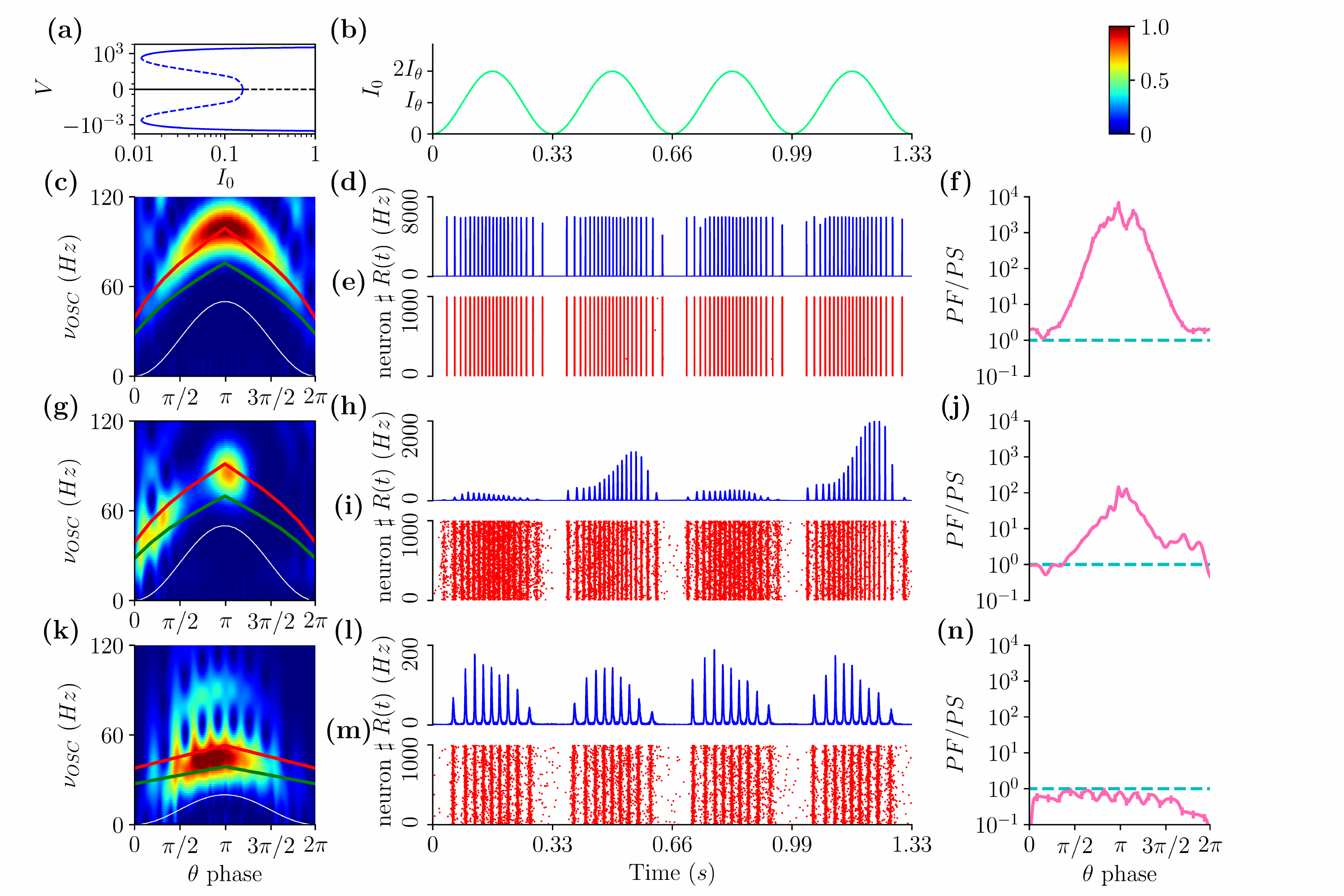}}
\caption{{\bf Fast and slow gamma oscillations entrainment with the theta forcing}
\red{
(a) Bifurcation diagram of the MF model analogous to the one reported in 
Fig. \ref{f15} (a). (b) Theta forcing \eqref{forcing} versus time.
The three lower rows refer from top to bottom to $I_\theta = 0.35$, 0.3 and 0.1.
In the left column are reported the normalized spectrograms as a function of the theta phase. 
In the same panels are reported $\nu_D$ (solid green line), 
$\nu_O$ (solid red line) as a function of $\theta$, as well as the forcing in arbitrary units (white solid line). The central column displays an instance over a short time
interval of the corresponding raster plots and instantaneous firing rates $R(t)$.
The right column reports the ratio $PF/PS$ of the power contained in the fast ($50< \nu_{OSC} < 100$ Hz) and slow
($30 \le \nu_{OSC} \le 50$ Hz) gamma bands as  a function of the $\theta$ phase.
In this case the error bars are displayed, but are almost invisible on the reported scale.
Parameters are $J_0=1$, $\tau_d = 0.15$ ms, $\Delta = 0.3$ and $K=1000$,
for the simulations we considered $N=10000$, $\nu_\theta =3$ Hz and $A_n= 1.1 \times 10^{-3}$,
the data for the spectrograms (left row) have been obtained by averaging over 30 theta cycles and
those for $PF/PS$ (right row) over 400 cycles.}
}
\label{f12}
\end{figure*}

\red{We considered the network dynamics in presence of the periodic forcing
\eqref{forcing} and additive noise on the membrane potentials (with zero mean and amplitude $A_n$).
As shown in Fig. \ref{f12}, the response of the system to the forcing is controlled by the value of the
amplitude $I_\theta$ in \eqref{forcing}: for small $I_\theta \le 0.20$ one observes
only slow gamma COs; for intermediate values of the amplitude $ 0.20 < I_\theta \le 0.32$
one has the coexistence of slow and fast gamma COs; while for $I_\theta \ge 0.32$ only
fast oscillations are present.
}

\red{For small $I_\theta$, as one can appreciate from the raster plot in panel (m),
the firings of the neurons, despite being partially synchronized, are quite irregular. 
Furthermore the corresponding spectrogram in Fig. \ref{f12} (k)
reveals that the power is concentrated at frequencies below 50 Hz and 
that the amplitude of the spectrum has a modulated structure as a function of the phase. 
This is confirmed by the analysis of the power of the spectrum $PS$ ($PL$) restricted 
to the slow  (fast) gamma band (see Fig. \ref{f12} (n)).
These are indications of theta-nested gamma oscillations, as confirmed by the
instantaneous firing rate reported in Fig. \ref{f12} (l), which reveals also
an evident P-A coupling between the gamma phases and the theta forcing.
}

\red{
These results resemble experimental observations of theta-nested gamma oscillations
induced {\it in vitro} by sinusoidal optical stimulation at theta frequency
in the medial entorhinal cortex (mEC) \cite{pastoll2013} and in the areas CA1 \cite{butler2016}
and CA3 \cite{akam2012} of the hippocampus. In all these experiments
single neurons spiked quite irregularly, while the collective dynamics
was oscillatory, analogously to our dynamics as shown in Fig. \ref{f12} (l) and (m).
As previously discussed, these COs are noise induced and characterized 
at a MF level by frequencies $\simeq \nu_D$ (green solid line), which
represents a reasonable
estimation of the position of the maxima of the spectrogram as shown in Fig. \ref{f12} (k).
}
 
\red{The situation is quite different for sufficiently large forcing amplitude,
where the neuronal dynamics becomes quite regular and almost fully synchronized,
as evident from Figs. \ref{f12} (d) and (e). In this case the power
is concentrated in the fast gamma band and it is maximal in correspondence
of the largest value of $I_0$ occuring at $\theta=\pi$ (see Figs. \ref{f12} (c) and (f)).
Furthermore the profile of the maximal power in the spectrogram 
follows reasonably well the MF values $\nu_O$ (red solid line) expected for
fast gamma COs, as evident from Fig. \ref{f12} (c).
For these large currents we have a sort of pathological synchronization
usually observable in connection with neuronal diseases. In particular,
highly synchronized fast gamma oscillations have been observed
in patients with neocortical epilepsy \cite{worrel2004}.
} 

\red{The most interesting situation occurs for intermediate amplitudes, specifically
we considered $I_\theta = 0.30$. As evident from Figs. \ref{f12} (h) and (i) in this case the network 
dynamics can vary noticeably from one theta cycle to the next,  due to the switching from 
one gamma regime to the other occurring erratically. However by averaging over a sufficiently 
large number of cycles we can identify stationary features of this dynamics. In particular, 
we observe that the values of maximum power in the spectrum correspond to different theta phases
for the slow and fast gamma COs: namely, for slow gamma the maximal activity is observable at small
angles, while for fast gamma this corresponds to the largest value 
of the forcing current \eqref{forcing} (see Figs. \ref{f12} (g) and (j)).
} 

\red{These findings are analogous to the experimental results reported in \cite{colgin2009}
for the region CA1 of the hippocampus in
freely moving rats, where it has been reported that slow gamma power were peaked
around $\theta \simeq 0.4 \pi$ and fast gamma power around $\theta \simeq \pi$,
corresponding also to the maximum of activity of excitatory place cells.
Similar results have been reported in \cite{belluscio2012} for what concerns
the slow gamma rhythm, however in those experiments fast gamma (referred in 
as intermediate gamma) occurs earlier in the theta cycle.}

\red{
The network response to the external periodic forcing \eqref{forcing}
can be interpreted  in terms of an adiabatic variation of the external current 
whenever the time scale of the forcing term is definitely slower with respect to
the neuronal time scales (i.e. $\tau_m$ and $\tau_d$). Since this is the case,
we can try to understand the observed dynamics at a first level of approximation 
by employng the bifurcation diagram of the MF model 
obtained for a constant DC current $I_0$, which is shown in Fig. \ref{f12} (a) 
for the set of parameters here considered. The diagram reveals that the system bifurcates
via a sub-critical Hopf from the asynchronous state to regular oscillatory behaviour at
a current $I^{(H)} \simeq 0.159$ and that the region of coexistence of stable foci and limit cycles is
delimited by a saddle-node bifurcation occuring at $I^{(S)} \simeq 0.012$
and by $I^{(H)}$.}

\red{The forcing current \eqref{forcing} varies over a theta cycle from 
a value $I_0 = 0$ at $\theta=0$ up to a maximal value $I_0 = 2 I_{\theta}$ at $\theta=\pi$
and returns to zero at $\theta = 2 \pi$. 
Since the forcing current will start from a zero value, we expect that the network will start oscillating with slow gamma frequencies associated to the stable focus which is the only stable solution at small $I_0 < I^{(S)}$. Furthermore, if $I_{\theta} < I^{(H)}/2$ the system will remain always in the slow gamma
regimes during the whole forcing period, since the focus is stable up to the current $I^{(H)}$.}

\red{ For amplitudes $I_{\theta} > I^{(H)}/2$ we 
expect a transition from slow to fast COs for a theta phase 
$\theta^{(H)} = \arccos{[(I_\theta-I^{(H)})/I_\theta]}$ corresponding to the
crossing of the sub-critical Hopf. Since this transition is histeretic 
the system will remain in the fast regime until the forcing current does not
become extremely small, namely $I_0 < I^{(S)}$, corresponding to
an theta phase $\theta^{(S)} = 2\pi - \arccos{[(I_\theta-I^{(S)})/I_\theta]}$.
}

\red{The performed analysis is quasi-static and does not take into account the time spent 
in each regime. If $I_\theta >> I^{(H)}$ the time spent by the system
in the slow gamma regime is extremely reduced, because $\theta^{(H)} \simeq 0$
and $\theta^{(S)} \simeq 2 \pi$, and this explains why for
large $I_\theta$ we essentially observe only fast gamma.
On the other hand, we find only slow gamma COs
for $I_\theta$ up to 0.20, a value definitely larger than $I^{(H)}/2$,
and this due to the fact that a finite crossing time is needed 
to jump from one state to the other as discussed in the previous 
sub-section.}

\red{
Let us now focus on the case $I_\theta=0.3$, where
we observe the coexistence of fast and slow gamma COs. 
As already mentioned we have stable foci in the range $I_0 \in [0 : I^{(H)}]$,
this in terms of $\theta$-angles obtained via
the relationship \eqref{forcing} for $I_\theta=0.3$ corresponds to an interval $\theta/\pi \in [0 : 0.34]$,
roughly matching the region of the spectrogram reported in Fig. \ref{f12} (g) where the maximum power of slow gamma oscillations is observable. As already mentioned, even if the forcing current $I_0(\theta)$
decreases for $\theta \to 2 \pi$, we would not observe slow gamma at large $\theta$-angles due to the histeretic
nature of the sub-critical Hopf transition.
Slow gamma COs are associated to fluctuation driven dynamics with frequencies $ \simeq \nu_D$ (green solid line), as confirmed also by the comparison with the maxima of the
power spectrum reported in Fig. \ref{f12} (g).
}

 \red{
For currents $I_0 > I^{(H)}$ only the limit cycles (corresponding to fast gamma COs with frequencies $\nu_O$)
are stable, indeed the maximum of the power spectrum for fast gamma COs occurs for $\theta \simeq \pi$
where $I_0 \simeq 0.6 > I^{(H)}$ is maximal. As expected, the CO frequency associated to the maximum of the power spectrum is well reproduced by $\nu_O$ (see the red solid line in Fig. \ref{f12} (g)).}

\begin{figure*}
\centerline{
\includegraphics[scale=0.9]{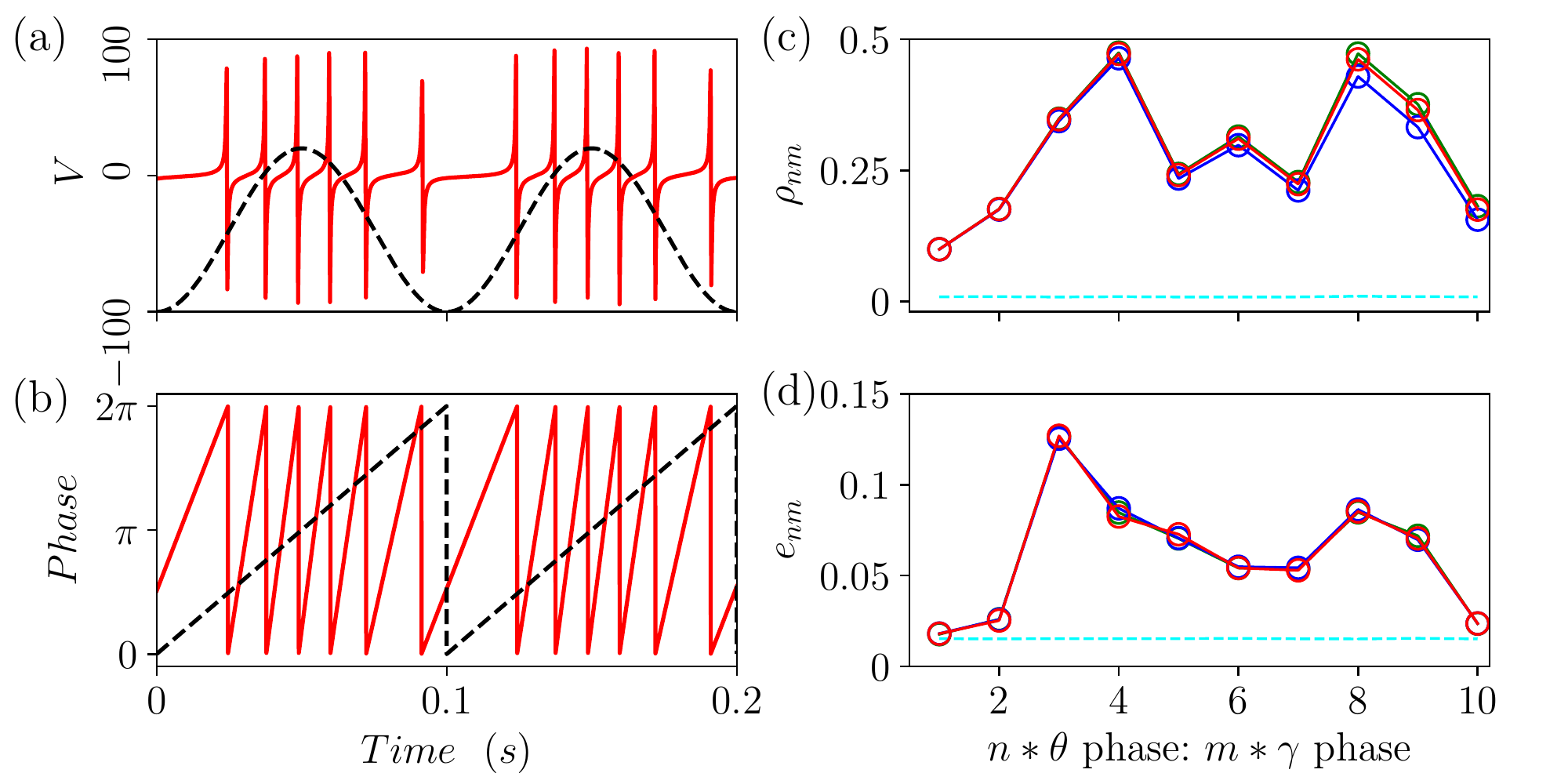}
}
\caption{{\bf Phase-phase coupling $n:m$ between theta forcing and gamma oscillations}
\red{(a-b) Locking of the gamma oscillations to the external theta forcing: (a)
average membrane potentila $V$ versus time, the black dashed line is the forcing \eqref{forcing}
in arbitrary units; (b) gamma (red solid) and theta (black dashed) phases for the corresponding
time interval. (c) Kuramoto order parameter $\rho_{nm}$ for the phase difference $\Delta_{nm}(t)$
for time windows of duration $T_W = 0.1$ s (black), $0.5$ s (red) and $1$ s (blue)
averaged over $70 < M < 700$ different realizations.
(d) Normalized entropy $e_{nm}$ for a time window $T_W=0.5$ s averaged over $M=140$ realizations (black),
surrogate data are also reported corresponding to random permutation (red) and time shift (blue) of
the original data averaged over $M=100$ independent realizations.
The reported data refer to the simulation of the
spiking network subject to the external forcing \eqref{forcing} with additive noise
on the membrane potentials. Parameters are  the same as in Fig. \ref{f12}, apart for 
$\nu_\theta =10$ Hz and $A_n= 1.0 \times 10^{-3}$, the histogram of $\Delta_{nm}(t)$ 
employed for the estimation of $e_{nm}$ have been evaluated over $M=50$ bins. 
The results refer only to phases associated to gamma frequencies in the band $30-100$ Hz. 
The the error bars in (c) and (d) are of the order of the size of the symbols and the
significance levels are
reported as dashed cyan lines in (b) $\rho^{(S)} = 0.009$ and (c) $e^{(S)} = 0.016$.}
}
\label{f13}
\end{figure*}

\red{
As a last point, let us examine if the coexistence of fast and
slow gamma COs is related
to some form of P-P locking between theta forcing
and gamma oscillations \cite{tass1998,belluscio2012}.
As evident from Fig. \ref{f13} (a) and (b) the theta
forcing at $\nu_\theta = 10$ Hz locks the collective network
dynamics, characterized by the mean membrane potential and by the $\gamma$-phase
defined in Section II C. In particular, for this specific time window
we observe for each $\theta$-oscillations exactly six $\gamma$-oscillations
of variable duration: slower at the extrema of the $\theta$-window and
faster in the central part. In agreement with the expected coexistence of $\gamma$ rhythms of
different frequencies.}

\red{Let us quantify these qualitative observations by considering statistical indicators 
measuring the level of $n:m$ synchronization
for irregular/noisy data over a large number of theta cycles. In particular, 
we will employ the Kuramoto order parameter $\rho_{nm}$  and the
normalized entropy $e_{nm}$ introduced in Section II C measured over time
windows of duration $T_W$ and averaged over many different realizations.}

\red{
As shown in Fig. \ref{f13} (c) and (d), both these indicators 
exhibit two maxima showing the existence of two different locking between $\theta$ and $\gamma$ oscillations for $n:m$ equal to $\simeq 3-4$ and $\simeq 8$, thus corresponding to slow and fast gamma (being $\nu_\theta = 10$ Hz). By following \cite{scheffer2016}, in order to test if the reported P-P couplings are significant,
we have estimated $\rho_{nm}$
over time windows of increasing duration, namely from 0.1 s to 1 s. As shown in 
Fig. \ref{f13} (c)  the measured values
do not vary substantially even by increasing $T_W$ by a factor $10$. This is a clear indication of the stationarity  of the P-P locking phenomenon here analysed \cite{scheffer2016}.
Furthermore, we measured  $e_{nm}$ also for surrogate data obtained by random permutation and by time-shift 
(for the exact definitions see Section II C and \cite{scheffer2016}), the values obtained
for these surrogate data are almost indistiguishable from the original ones (see Fig. \ref{f13} (d)).
These results demand for the development of more effective approaches
able to distinguish true locked state from spurious locking.
}

\red{
Fnally, the significance level of the reported measurements have been evaluated
by randomly shuffling the time stamps of the $\gamma$-phases and 
 denoted as $\rho^{(S)}$ and $e^{(S)}$, respectively
(dashed lines in  Fig. \ref{f13} (c) an (d)).
The values of $\rho^{(S)}$ and $e^{(S)}$ are definitely smaller
than those of the corresponding indicators in correspondence 
of the observed P-P lockings,
thus confirming their significance.}

%figure 10

%figure 11

\section{Conclusions}

In this paper we have shown in terms of an effective 
mean-field that in a sparse balanced inhibitory network
with finite synaptic decay COs can emerge via
super or sub-critical Hopf bifurcations from 
a stable focus. Furthermore, in the network (for sufficiently low
structural heterogeneity) the macroscopic focus turns out
to be unstable towards microscopic fluctuations in the
firing activity leading to the emergence of COs
characterized by a frequency corresponding to that
of the damped oscillations towards the MF focus.
Therefore in proximity of the sub-critical Hopf bifurcations
the coexistence of two COs with different origins is observable:
slow (fast) gamma oscillations being fluctuation (mean) driven.

\red{From our analysis it emerges that two ingredients
are needed to observe coexisting slow and fast gamma COs:
the sparsness in the connections and the dynamical balance
of the network activity. In particular, the sparsness 
has a twofold effect at the macroscopic and at the microscopic level.
In a mean-field framework the randomness in the in-degree distribution
can be reinterpreted as a quenched disorder in the synaptic couplings, 
which gives rise to the coexistence of stable foci and limit cycles.
However, in a fully coupled network with heterogeneous parameters 
we would not observe strong irregular fluctuations at the level of single neurons,
 analogous to Poissonian-like firings ususally observed in the cortex
 \cite{ullner2016,angulo2017,devalle2017}. These can emerge only in sparsely
 connected networks \cite{brunel1999,brunel2000}. 
Moreover, the balance mechanism guarantees 
that the irregular spiking dynamics will not disappear
in the thermodynamic limit \cite{bal1,bal2,bal3,matteo}.
These persistent microscopic fluctuations are 
able to trigger slow gamma COs in the network, which coexist with fast gamma 
COs corresponding to the limit cycle solutions in the MF.
These two ingredients usually characterize real brain networks,} where 
our prediction that slow (fast) gamma oscillations are associated to more (less) irregular neuronal dynamics can
be experimentally tested. e.g. by measuring the coefficient of
variation associated to these two states.  
Furthermore, previous theoretical analysis of gamma oscillations based on two interacting Wilson-Cowan rate models with different synaptic times revealed only the possible coexistence of two stable limit cycles both corresponding to tonic collective firing (i.e. mean driven COs) \cite{keeley2016}.

\red{Our model is not meant to explicitly replicate 
the dynamics of specific brain areas, but rather to 
illustrate fundamental mechanisms by which slow and fast gamma oscillations
may arise and coexist due to local network inhibitory features.
However, several phenomena we reported resemble experimental
results obtained for different brain regions {\it in vitro}
as well as {\it in vivo} and our findings can stimulate new experiments
or lead to novel interpretation of the existing data}.

\red{Of particular interest is the possibility, analysed in
Section V A, to switch from a gamma rhythm to the other by performing 
transient stimulations. This mechanism can allow
a single inhibitory population to pass from a coding task
to another following an external sensory stimulus. Indeed
it has been shown that distinct gamma rhythms are involved
in different coding processes: namely, fast gamma in new memory encoding, 
while slow gamma has been hypothized to promote memory retrieval \cite{mably2018}. 
}

\red{ On one side, pathological synchronization is usually associated to neuronal
diseases \cite{hammond2007,oswal2013,truccolo2014}. On another side, aberrant gamma oscillations have been
observed in several cognitive disorders, including Alzheimer’s
disease, Fragile X syndrome and neocortical epilepsy \cite{worrel2004,mably2018}.
Furthermore, deep brain stimulation (DBS) techniques have been developed along the
years to treat some of these diseases, e.g. essential tremor and Parkinsons´ disease   \cite{perlmutter2006,breit2004,de2015}. We have presented a simple model exhibiting the coexistence of
highly synchronized states and asynchronous or partially synchronized
regimes. Therefore, our model can represent a simple benchmark where to test 
new DBS protocols to obtain eventually less invasive technique to
desynchronize pathological states \cite{tass2002,popovych2005,andrews2010}
or to restore healthy gamma rhythms, as suggested in \cite{mably2018}.}

Moreover, the richness of the dynamical scenario present in this simple model indicates possible future directions where intrinsic mechanisms present in real neural networks like spike frequency adaptation could permit a dynamical alternation between different states. In this direction, a slow variable like adaptation could drive the system from "healthy" asynchronous or oscillatory dynamics to  periods of pathological extremely synchronous regimes, somehow similar to epileptic seizure dynamics \cite{jirsa2014}.

\red{In Section V B, we have analysed the emergence of COs in our network
in presence of an external theta forcing. This in order to 
to make a closer contact with recent experimental investigations
devoted to analyse the emergence of gamma oscillations in several
brain areas {\it in vitro} under sinusoidally modulated theta-frequency
optogenetic stimulations \cite{akam2012,pastoll2013,butler2016}.
For low forcing amplitudes, our network model displays theta-nested gamma COs
at frequencies $\simeq 50$ Hz joined with irregular spiking dynamics. 
These results are analogous to the ones reported
for the CA1 and CA3 areas of the hippocampus in \cite{akam2012,butler2016},
moreover theta-nested oscillations with similar features have been reported also 
for the mEC \cite{pastoll2013}, but for higher gamma frequencies.
}

\red{Furthermore, for intermediate forcing amplitudes
we observe the coexistence of slow and fast gamma oscillations,
which lock to different phases of the theta rhythm,
analogously to what reported for the rat hippocampus during exploration and REM sleep 
\cite{colgin2009, belluscio2012}. The theta-phases preferences displayed
in our model by the different gamma rhythms are due to the histeretic 
nature of the sub-critical Hopf bifurcation crossed
during the theta stimulation. 
Finally, for sufficiently strong forcing, the model is driven
in the fast gamma regime.}

\red{Our analysis suggests that a single inhibitory
population can generate locally different gamma rhythms
and lock to one or the other in presence of a theta forcing. 
In particular, we have shown that fast gamma oscillations are locked to a strong excitatory input, 
while slow gamma COs emerge when excitation and inhibition balance. 
These results can be useful in revealing the mechanism behind
slow and fast gamma oscillations reported in several brain
areas: namely, hippocampus \cite{colgin2016},
olfactory bulb \cite{kay2003}, ventral striatum \cite{van2009},
visual cortical areas \cite{bastos2015} and neocortex \cite{sirota2008}.
Particularly interesting are the clear evidences
reported in \cite{sirota2008} that different gamma rhythms,
phase locked to the hippocampal theta rhythm, can be 
locally generated in the neocortex. Therefore future
studies could focus on this brain region to test 
for the validity of the mechanisms here reported.}

\red{For what concerns the CA1 area of the hippocampus,
where most of the experimental studies on theta-gamma oscillations
have been performed. Despite the experimental evidences that different
gamma oscillations emerging in CA1 area at different theta
phases are a reflection of synaptic inputs originating from CA3 area 
and mEC \cite{colgin2009,schomburg2014} this does not
exclude the possibility that a single CA1 inhibitory population
can give rise to different gamma rhythms depending on the 
network state \cite{colgin2016}. This hypothese is supported 
by experimental evidences showing that a large part of CA1 interneurons {\it in vivo} can lock to
both slow and fast gamma \cite{colgin2009,belluscio2012,schomburg2014}
and that {\it in vitro} gamma rhythms can be locally generated in various
regions of the hippocampus due to optogenetic stimulations
\cite{akam2012,pastoll2013,butler2016}
or pharmalogical manipulation \cite{fisahn1998,traub2003,pietersen2014,craig2015}.
However, much work remains to be done to clarify
if local mechanisms can give rise to coexisting gamma rhythms also in the CA1 area.}

\red{At variance with previous results for purely inhibitory populations
reporting noise sustained COs in the range $100-200$ Hz \cite{buzsaki2012}
our model displays slow gamma rhythms characterized by irregular firing of the single neurons.
Therefore in our case it is not necessary to add an excitatory population to the inhibitory
one to slow down the rhythm and to obtain oscillations in the gamma range as done in \cite{brunel2003,geisler2005}. Evidences have been recently reported pointing out that 
gamma oscillations can emerge locally in the CA1 induced by the application of kainate
due to purely inhibitory mechanisms \cite{craig2015}. However, other studies point out 
that in the same area of the hippocampus excitatory and inhibitory neurons should 
interact to give rise to oscillations in the gamma range \cite{pietersen2014,butler2016}.
Preliminary results obtained for QIF networks with a sinusoidal
theta forcing show that theta-nested gamma oscillations 
with similar features can emerge for purely inhibitory as well as for mixed excitatory-inhibitory networks \cite{marco}.
}

\red{As shown in Section III B the same kind of bifurcation diagram can be
observed by considering the external excitatory drive as well as the self-disinhibition
of the recurrently coupled inhibitory population.
This suggests that in our model the same scenarios reported in Section V 
for an excitatory theta forcing can be obtained by considering an external inhibitory population 
which transmits rhythmically its activity to the target population. This somehow mimicks the
pacemaker theta activity of a part of the medial septum interneurons on the interneurons 
of the hippocampus experimentally observed in \cite{hangya2009}. This subject will be
addressed in future studies due to its relevance in order to clarify the origin of 
theta-gamma oscillations in the hippocampus, however it goes beyond the scopes of the present analysis.
}

In this paper we considered a model including the minimal ingredients 
necessary to reproduce the phenomenon of coexisting gamma oscillations
corresponding to quite simple (namely, periodic) collective regimes.
However, the introduction of synaptic delay in the model can lead to more
complex coexisting states, like quasi-periodic and even chaotic solutions, 
as recently shown for fully coupled networks in \cite{pazo2016,devalle2018}.
The inclusion of delay in our model
can enrich the dynamical scenario maybe allowing to mimic further aspects of
the complex patterns of activity observed in the brain, like e.g.
sharp-wave ripples observed in the hippocampus and
which are fundamental for memory consolidation 
\cite{buzsaki2015}. \red{Due to the large variety of interneurons 
present in the brain and in particular in the hippocampus \cite{maccaferri2003}
a further step in rendering our model more realistic would consist in considering
multiple inhibitory populations characterized by different neuronal parameters.
By manipulating the influence of a population on the others it would be
interesting to investigate the possible mechanisms to switch COs from one gamma rhythm to another, 
following the richness of the bifurcation scenarios presented in Figs. 2 and 3.}

\red{The generality of the phenomena here reported is addressed in Appendix A and B.
In particular in Appendix A we show that the mechanisms leading to the 
coexistence scenario of fast and slow gamma
oscillations are not peculiar of Lorentzian in-degree distributions (that we employed
to allow a comparison of the network simulations with the MF results), but that they are observable
also in the more studied Erd\"os-Reniy sparse networks. Appendix B is devoted
to the analysis of a  suitable normal form, which reproduces the dynamics of the MF
in proximity of the sub-critical Hopf bifurcation.  
In particular, the noisy dynamics of the normal form reveals coexisting
oscillations of different frequencies.
More specifically the  addition of noise leads from damped oscillations towards 
the stable focus to sustained oscillations characterized by the same frequency.
This latter result links our findings to 
the more general context of noise-induce oscillations for non-excitable systems 
examined in various fields of research: namely, single cell oscillations 
\cite{thomas2013}, epidemics \cite{kuske2007}, predator-prey interactions \cite{rozenfeld2001}
and laser dynamics \cite{ushakov2005}. At variance with all previous studies
we have analyzed noise-induced oscillations coexisting and interacting with oscillations emerging
from the Hopf bifurcation. Furthermore, the mechanism leading to the irregular fluctuations
in our case is quite peculiar. Single cells oscillations
are believed to be driven by molecular noise, induced by the small number of molecules
present in each cell, and therefore disappearing in the thermodynamic limit \cite{vilar2002}.  Recently, another possible mechanism leading to fluctuation amplification in a feed-forward chain has been suggested as a pacemaking mechanism for biological systems,
in this context the amplitude of the oscillations grows with the system size  \cite{fanelli2017}. 
Instead in our case, the dynamical balance provides intrinsic noise and oscillations
of constant amplitude, essentially independent from the number of synaptic inputs (in-degree) 
and from the number of neurons in the network.  
}

\acknowledgements
We acknowledge useful discussions with D. Avitabile, D. Angulo-Garcia, F. Devalle, S. Keeley, E. Montbri\'o, S. Olmi, A. Politi, J. Rinzel, and R. Schmidt. The authors received partial economic support by the French Governement via the Excellence  Initiative I-Site Paris Seine (Grant No ANR-16-IDEX-008), the Labex MME-DII (Grant No ANR-11-LBX-0023-01) and the ANR Project ERMUNDY (Grant No 18-CE37-0014-03).

\appendix
\section{Slow and fast gamma oscillations in Erd{\"o}s-Reniy network }

\begin{figure}
\centerline{\includegraphics[scale=0.35]{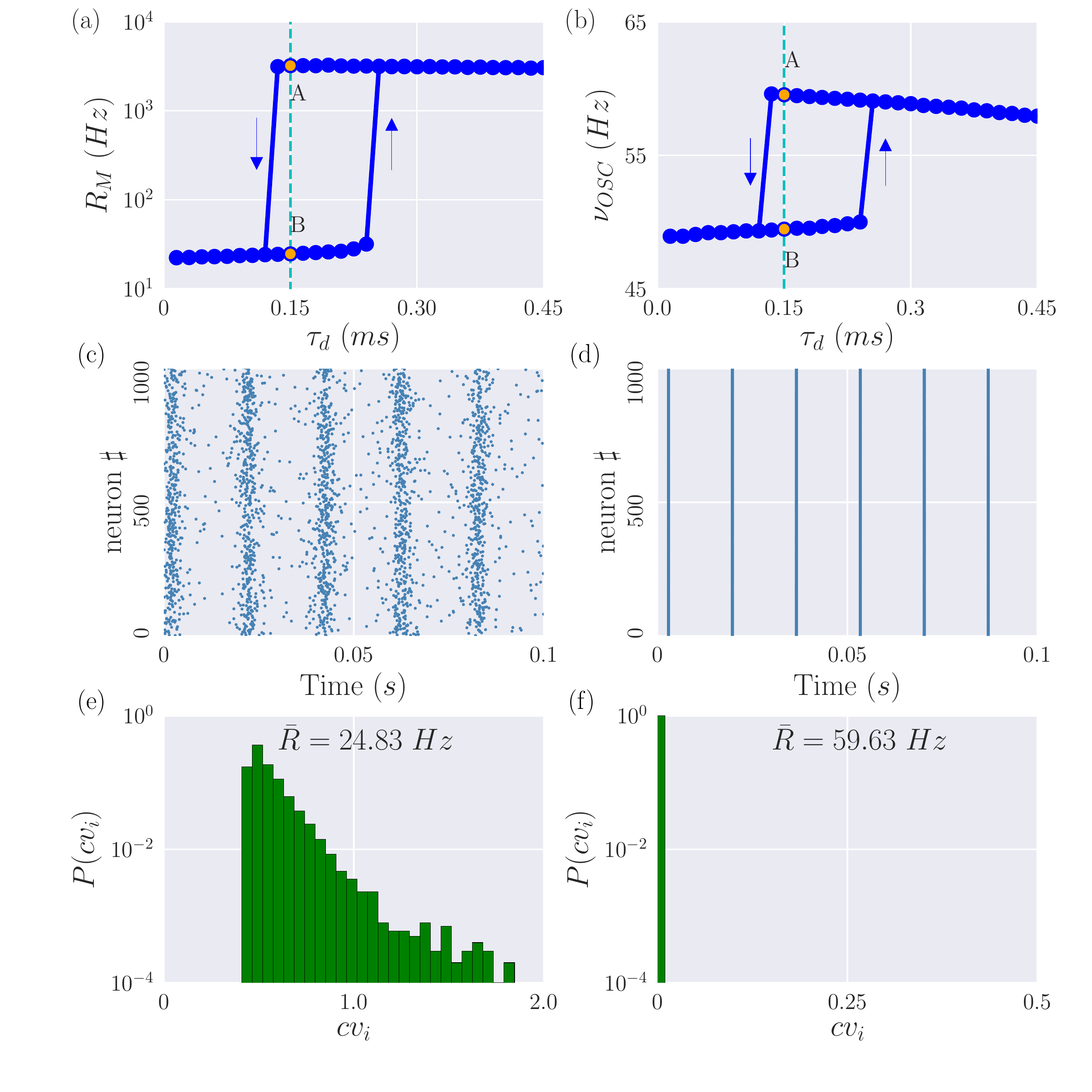}}
\caption{{\bf Erd\"os-Reniy Network}
Results of adiabatic simulations for an ER network obtained
by varying the synaptic time $\tau_d$:
(a) maximal firing rates ${R_M}$ and (b) frequencies $\nu_{OSC}$ of the
COs. Two coexisting states (A) and (B) are considered at $\tau_d = 0.15$ ms.
In the left and right row are reported the raster plots (c,d) and the distributions
of the $cv_i$ (e,f) for the state (A) and (B), respectively.
Parameters for the simulations are $N=10000$, $K=1000$, $I_0 = 0.25$, $J_0 = 1.0$
and $\Delta \tau_d = 0.015$ ms, $\tau_d^{(0)} = 0.015 $ ms, $\tau_d^{(1)} = 0.45$ ms.
}
\label{f8}
\end{figure}

In order to compare the network simulations with
the MF results we have considered in the article a Lorentzian
distribution for the in-degrees. It is therefore important to
show that the same phenomenology is observable by considering
a more standard distribution, like the Erd\"os-Reniy (ER) one.
The results of adiabatic simulations, reported in Fig. \ref{f8}, confirm that also
for ER networks a bistable regime, characterized 
by COs with different gamma-frequencies (see panel (b)), is indeed observable.
In particular, slow gamma COs characterized by an average firing rate ${\bar R} \simeq 25 $ Hz
and irregular neuronal firings (as shown in panels (c) and (e))
coexist with almost synchronized fast gamma COs with neurons tonically firing with
${\bar R} \simeq 60 $ Hz (see panels (d) and (f)).

%figure 12

\section{A general mechanism for the emergence of coexisting oscillations}

\red{
We investigate here the generality of the mechanism for the coexistence of COs 
observed in the network of QIF neurons. 
In particular, we have shown that this phenomenon occurs when in the MF model we have a
focus coexisting with a limit cycle, while in the sparse network we have fluctuations sustained
by the dynamical balance. If this is the mechanism we expect to see
a similar phenomenon whenever we consider a system in proximity of 
a sub-critical Hopf bifurcation and we add noise of constant amplitude to the dynamics.}

\red{Therefore, to asses the generality of the phenomenon we consider the normal form of a Hopf bifurcation
in two dimensions leading to the birth of a limit cycle from an equilibrium, namely  \cite{andronov1971theory,kuznetsov2013elements}:
\begin{eqnarray}
& \hspace*{-1cm} \tau_m \dot x =  \beta x - y +\sigma x r^2-(x+\gamma y)r^4+I_1
%%\label{normal_y1}
\\
& \hspace*{-0.8cm}  \tau_m \dot y =   x + \beta y +\sigma yr^2+(\gamma x-y)r^4+I_2\enskip,
\label{normal_f}
\end{eqnarray}
where $r^2=x^2+y^2$, $\tau_m =4$ ms is an arbitrary time scale, $I_{1}(t)$ and $I_2(t)$ are generic external time dependent forcing, $\beta$ is the bifurcation parameter, the parameter $\sigma$ sets the nature of the bifurcation and $\gamma$ controls the frequency of the stable and unstable limit cycles.
Notice that we added a quintic term, absent in the original normal form  \cite{andronov1971theory,kuznetsov2013elements}, in order to 
maintain bounded the values of  $x$ and $y$ while keeping the same bifurcation structure.
For $I_1=I_2=0$ we will have a sub-critical (super-critical) Hopf for $\sigma=+1$ ($\sigma=-1$). In this case
it is convenient to rewrite \eqref{normal_f} in polar coordinates $(x,y)=(r\cos{\phi},r\sin{\phi})$,
as follows:
\begin{eqnarray}
\tau_m \dot r &=&  \beta r + \sigma r^3 - r^5
%%\label{normal_y1}
\\
\tau_m \dot \phi &=& 1 + \gamma r^4 \quad .
\label{normal_polar}
\end{eqnarray}
The stationary solutions are $r=0$ corresponding to stable focus characterized by relaxation
oscillations with a frequency $\nu_{D} \simeq 39$ Hz and a stable and unstable limit cycles of
amplitudes $r^2=(\sigma \pm \sqrt{\sigma^2 + 4 \beta})/2$. 
}

\begin{figure}
\begin{centering}
\includegraphics[width=0.5\textwidth,clip=true]{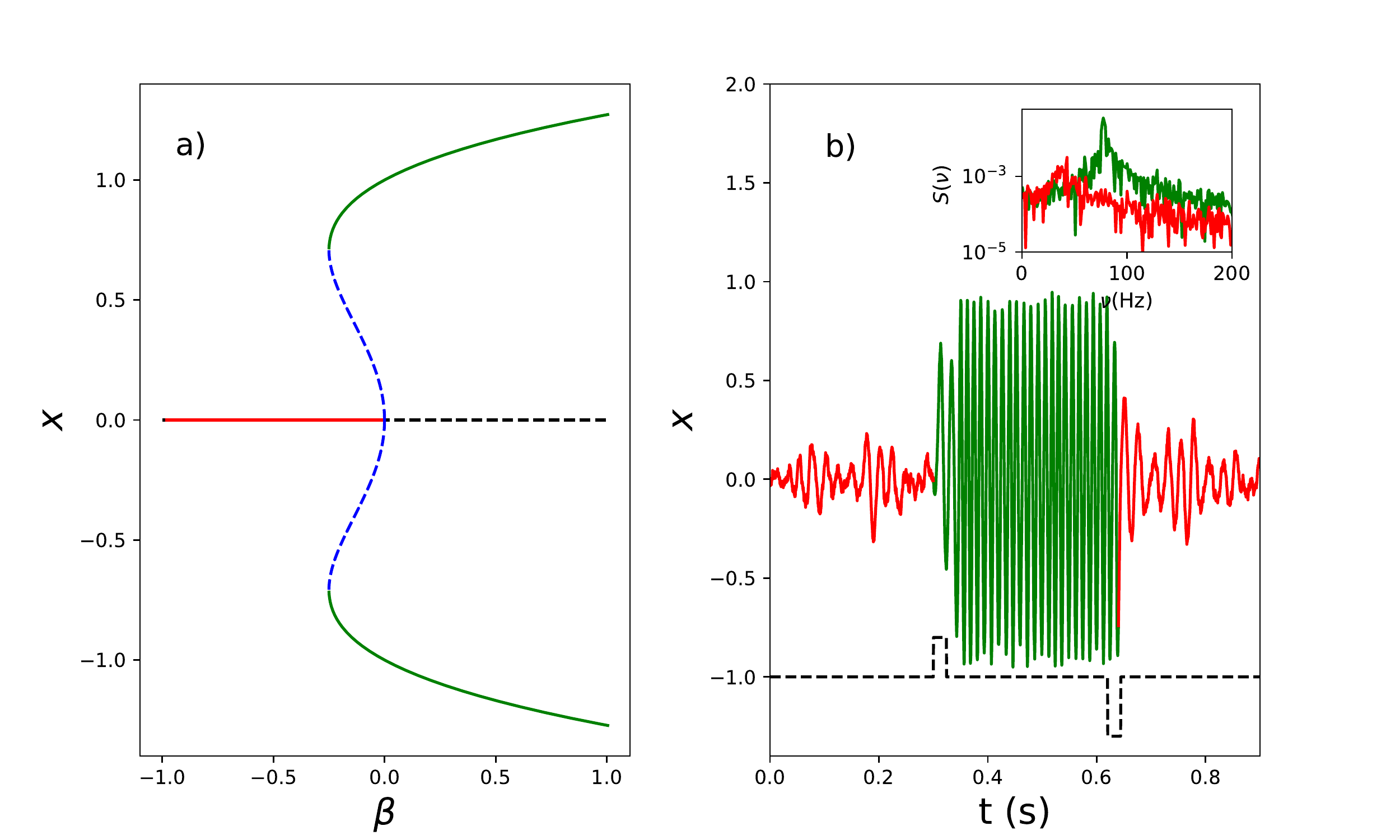}
\end{centering}
\caption{\label{fig:normal}
\red{
a) Bifurcation diagram for the variable $x$ as a function of the parameter $\beta$. Green (blue) lines indicate a stable (unstable) limit cycle and red (black) line a stable (unstable) focus. b) Fixing $\beta$ in the bistability region we report the time trace of $x(t)$ in presence of a zero-mean gaussian noise of amplitude $A_1=A_2=0.14$. An external pulse of current is added to the evolution equation for $x$ in
\eqref{normal_f} for a time window of $56$ ms to induce a switching between the oscillatory states (the black dashed line, shifted on the $x$
axe to be visible while the actual baseline value is zero). In the inset we report the power spectrum of the two different oscillatory regimes obtained over long time traces (hundreds of seconds) in order to check that the oscillations persist in time. Parameters are $\beta=-0.16$, $\sigma=1$, $\gamma=1.5$.}}
\end{figure}

\red{
In Fig. \ref{fig:normal} (a) we report the bifurcation diagram for $\sigma=+1$  and $I_1=I_2=0$. We observe that the sub-critical Hopf bifurcation occurs at $\beta=\beta_c=0$ and for $\beta < 0$ it exists a region where a stable (green dots) and unstable (blue dashed line) limit cycles coexists with a stable focus (red line), exactly as it happens for the QIF MF model (see Fig. \ref{f4} (a)). The stable and unstable limit cycles merge at a SN bifurcation located at $\beta = - \sigma^2/4$.
} 

\red{
As previously stated, the MF model cannot capture the endogenous fluctuations, naturally present in  
sparse balanced networks. In order to emulate this effect we consider $I_1(t)$ and $I_2(t)$ to be  two i.i.d. Gaussian white noise processes (i.e. $I_q(t) = A_q \xi_q(t) $ with $q=1,2$, where $\xi_q(t)$ are random, Gaussian distributed, variables of zero average and unitary variance). 
In presence of these additive noise terms and in proximity of the Hopf bifurcation, we observe the coexistence of two oscillatory regimes  as shown in Fig. \ref{fig:normal} (b).}
\red{One oscillation, characterized by higher amplitude (green line), 
corresponds to the limit cycle present in the non-noisy dynamics (green line in the bifurcation diagram reported 
in  Fig. \ref{fig:normal} (b).). The other oscillation is the result of a constructive role of noise that excites the stable focus thus generating robust oscillations at the frequency $\nu_D$ (red line). Analogously
to what shown for the network of QIF neurons (see Fig.  \ref{f9}), it is possible to switch between the two kind of oscillations via a pulse current of positive (negative) amplitude with respect to the baseline (see the dashed line in panel b)). Moreover the frequencies of the two oscillations, generated by two different mechanisms, corresponds
to slow and fast gamma oscillations as observable in the corresponding power spectra $S(\nu)$
reported in the inset of Fig. \ref{fig:normal} (b).
.}

%%\bibliographystyle{apsrev-nourl}
%%%\bibliographystyle{plain}

%%\bibliography{../lif_ost_cd}

\begin{thebibliography}{107}
\bibitem[{fie()}]{field}
\bibinfo{note}{In the definiton of the mean synaptic field we have divided the
  sum also by the median in-degree $K$, because on average a neuron is subject
  to $K$ spike trains.}
\end{thebibliography}

\end{document}